\documentclass[10pt,journal,compsoc]{IEEEtran}

\ifCLASSINFOpdf
\else
\fi
\usepackage{amssymb}
\usepackage{pifont}
\usepackage{multicol}
\usepackage{epsfig}
\usepackage{url}

\usepackage{hyperref}

\usepackage{multirow}
\usepackage{comment}
\usepackage{float}
\usepackage{varioref}
\usepackage{listings}
\usepackage{color}

\usepackage{graphicx}
\graphicspath{{./figures/}}
\usepackage{subcaption}
\usepackage[color = green!50]{todonotes}
\usepackage{amsmath}
\usepackage{cleveref}
\usepackage{algpseudocode}
\usepackage[linesnumbered,ruled]{algorithm2e}
 \usepackage{longtable}
\usepackage{diagbox}
 \usepackage{multirow}
\usepackage{booktabs} 
 \usepackage{algpseudocode}
 \usepackage[linesnumbered,ruled]{algorithm2e}
 \usepackage[utf8]{inputenc}
 \usepackage{colortbl}
 \usepackage{listings}
\usepackage{color}
 \usepackage{amsmath}
 \usepackage{titlesec}
\usepackage{array,tabularx}
\usepackage{makecell}
\usepackage[thinc]{esdiff}

\newenvironment{conditions*}
  {\par\vspace*{\abovedisplayskip}\noindent
   \tabularx{\columnwidth}{>{$}l<{$} @{${}={}$} >{\raggedright\arraybackslash}X}}
  {\endtabularx\par\vspace*{\belowdisplayskip}}
  
\usepackage{graphicx}

\lstdefinelanguage
   [sparc]{Assembler}     
   [x86masm]{Assembler} 
   {
   alsoletter={0123456789},
       keywords=[0]{
       be,nop,ld,cmp,if,mov,srl,xor,and,bne
   },
   keywords=[1]{g0,g1,g2,g3,g4,g5,g6,fp},
   keywords=[2]{0,1,2,3,4,5,6,7,8,9,10,11,12,13,14,15,16,17,18,19,20,24}
   }

\definecolor{dkgreen}{rgb}{0,0.52,0}
\definecolor{lightgray}{rgb}{0.97,0.97,0.97}
\definecolor{gray}{rgb}{0.44,0.44,0.44}
\definecolor{dkgray}{rgb}{0.24,0.24,0.24}
\definecolor{violet}{rgb}{0.58,0,0.82}
\definecolor{red}{rgb}{0.58,0,0.0}

\usepackage{adjustbox}
\usepackage{array}

\newcolumntype{R}[2]{%
    >{\adjustbox{angle=#1,lap=\width-(#2)}\bgroup}%
    l%
    <{\egroup}%
}

\usepackage[printonlyused]{acronym}
\newacro{pdn}  [PDN]  {Power Distribution Network}
\newacro{ro}  [RO]  {Ring Oscillator}
\newacro{pro}  [PRO]  {Programmable Ring Oscillator}
\newacro{emfi}  [EMFI]  {Electromagnetic Fault Injection}
\newacro{sca}  [SCA]  {Side-Channel Analysis}
\newacro{tdc}  [TDC]  {Time-to-Digital Converter}
\newacro{cpa}  [CPA]  {Correlation Power Analysis}
\newacro{dpa}  [DPA]  {Differential Power Analysis}
\newacro{spa}  [SPA]  {Simple Power Analysis}
\newacro{snr}  [SNR]  {Signal-to-Noise Ratio}
\newacro{pll}  [PLL]  {Phase-Locked Loop}
\newacro{soc}  [SoC]  {System-on-Chip}
\newacro{prng}  [PRNG]  {Pseudo-Random Number Generator}

\begin{document}
%
\title{Programmable RO (PRO): A Multipurpose Countermeasure against Side-channel and Fault Injection Attacks} 
\title{Programmable Ring Oscillator (PRO): A Multipurpose Design for On-chip Side-channel Countermeasure and Fault Detection}

\author{Yuan~Yao,~\IEEEmembership{Student Member,~IEEE,}
        Pantea~Kiaei,~\IEEEmembership{Student Member,~IEEE,}
        Richa~Singh,
        Shahin~Tajik,
        and~Patrick~Schaumont,~\IEEEmembership{Senior~Member,~IEEE}
\IEEEcompsocitemizethanks{\IEEEcompsocthanksitem Y. Yao was with the Bradley Department
of Electrical and Computer Engineering, Virginia Polytechnique Institute and State University, Blacksburg,
VA, 24061.\protect E-mail: yuan9@vt.edu \\
\IEEEcompsocthanksitem P. Kiaei, R. Singh, S. Tajik, and P. Schaumont were with the Department
of Electrical and Computer Engineering, Worcester Polytechnique Institute, Worcester,
MA, 01609.\protect E-mail: \{pkiaei, rsingh7, stajik, pschaumont\}@wpi.edu \\

}
\thanks{This research was funded in part by National Science Foundation (NSF) grant 1931639.}

}

\IEEEtitleabstractindextext{%
\begin{abstract}
Side-channel and fault injection attacks reveal secret information by monitoring or manipulating the physical effects of computations involving secret variables. Circuit-level countermeasures help to deter these attacks, and traditionally such countermeasures have been developed for each attack vector separately. We demonstrate a multipurpose ring oscillator design - Programmable Ring Oscillator (PRO) to address both fault attacks and side-channel attacks in a generic, application-independent manner. PRO, as an integrated primitive,  can provide on-chip side-channel resistance, power monitoring, and fault detection capabilities to a secure design. We present a grid of PROs monitoring the on-chip power network to detect anomalies. Such power anomalies may be caused by external factors such as electromagnetic fault injection and power glitches, as well as by internal factors such as hardware Trojans. By monitoring the frequency of the ring oscillators, we are able to detect the on-chip power anomaly in time as well as in location.
Moreover, we show that the PROs can also inject a random noise pattern into a design's power consumption. By randomly switching the frequency of a ring oscillator, the resulting power-noise pattern significantly reduces the power-based side-channel leakage of a cipher. We discuss the design of PRO and present measurement results on a Xilinx Spartan-6 FPGA prototype, and we show that side-channel and fault vulnerabilities can be addressed at a low cost by introducing PRO to the design. We conclude that PRO can serve as an application-independent, multipurpose countermeasure. 
\end{abstract}
}

\maketitle

\section{Introduction}


In a physical side-channel attack, an adversary learns secret information by passively monitoring or else actively influencing the implementation of a secure electronic system. While power consumption is a popular target in side-channel attacks, many other sources of physical quantities have been identified and used as side-channel leakage. Besides passive monitoring of circuit behavior, an additional cause of information leakage stems from targeted faults. By analyzing the corresponding fault response, an attacker can retrieve the secret information from a target \cite{bar2006sorcerer}. The most common methods to inject faults include power glitches, clock glitches, electromagnetic pulses, and laser pulses. Finally, fault injection and side-channel monitoring can also be used in a combined attack, for example, to break a masking side-channel countermeasure \cite{yao2018fault}. 




%
Even though many existing works have demonstrated side-channel and fault attack countermeasures, there are no simple circuit-level solutions to solve \textbf{both} side-channel and fault attack vulnerabilities in a generic manner.
Generally, even for individual side-channel or fault countermeasures, a significant overhead will be introduced to the design. Moreover, many of the existing countermeasure mechanisms have to be specifically adjusted for the implemented algorithm.
\begin{figure}
    \centering
    \includegraphics[width=1.1\linewidth]{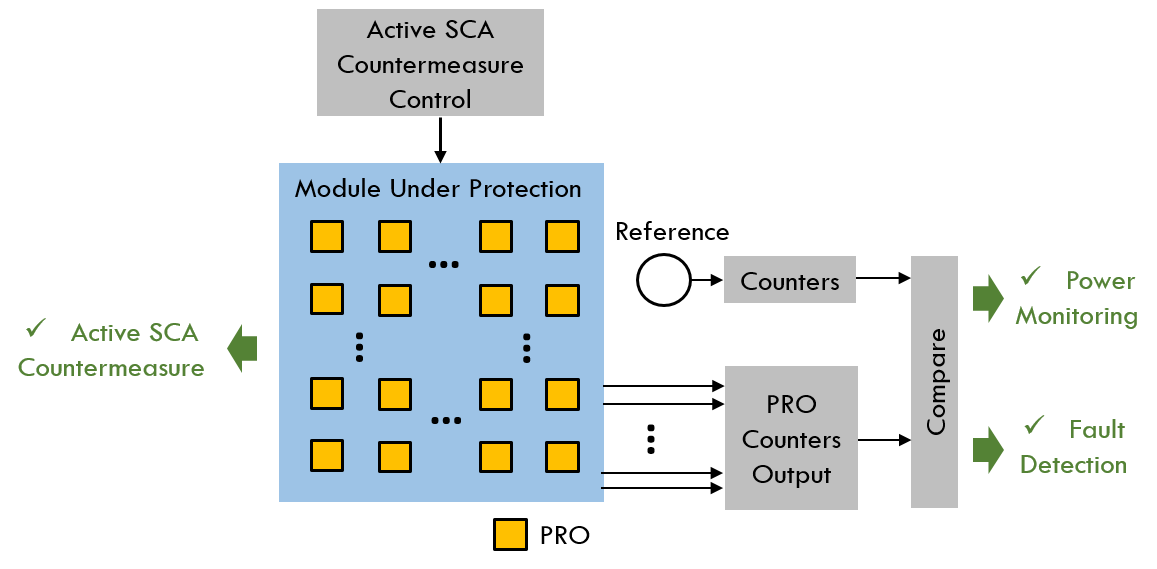}
    \caption{PRO based on-chip Secure Network Hardware Extension}
    \label{fig:intro}
\end{figure}

In recent years, researchers have further demonstrated that the placement of the attacker and the victim circuitry on the same chip while sharing a common \ac{pdn} brings new side-channel and fault attack opportunities. Having a common \ac{pdn} intrinsically relates the perturbations from the victim's logic to the attacker's logic and vice versa. Therefore, a neighboring adversary logic can interpret information about the victim operations by monitoring the changes on the shared \ac{pdn}. On the other hand, the same physical effect exists in the other way around; The victim logic can infer malicious operations of its neighbor circuitry by monitoring the shared \ac{pdn}. Therefore, in order to guarantee the security of the \ac{pdn}, a monitoring sensor network on the \ac{pdn} should be built to detect on-going attacks. The monitoring sensor network should fullfill the requirements including large spatial coverage, i.e., covering the full \ac{pdn} area, and large temporal coverage, i.e., continuously monitoring the \ac{pdn} \cite{tajik2017pufmon}. 

Previously, \acp{ro} were widely used by silicon design houses as test structures or on-chip sensors to monitor the performance of their technologies and circuits \cite{kim2008cmos}. But a multi-purpose design of RO-based on-chip sensors has not been investigated in adding resistance against both side-channel and fault attacks to the circuit. 
In this work, we introduce a new multi-purpose Ring Oscillator design - programmable RO (PRO). With a low overhead, the proposed PRO can provide the following solutions within the same structure: 
\begin{itemize}
    \item Active Side-channel hiding countermeasure;
    \item On-chip power monitoring;
    \item Fault injection monitoring;
\end{itemize}

The proposed PRO design has multiple configurations of oscillation frequency, which are under the control of the user (i.e., the defender). 
Each PRO has its own counter which can be read to calculate the PRO's frequency by comparing it with a reference counter. We first demonstrate that with low overhead, an individual PRO can provide sufficient disturbance to the power to hide the side-channel leakage of the secret information in the system. Moreover, we further demonstrate that by combining multiple PROs into an array and by placing them within the module under protection, a secure on-chip monitoring network can be constructed to monitor the power fluctuations on the \ac{pdn} to detect abnormalities and fault attacks. 

\autoref{fig:intro} shows the overall structure of the PRO-based on-chip secure system. The PROs are evenly placed on the chip to form a secure on-chip network. The PRO secure network can be controlled by the external user configuration. The user can turn on the \ac{sca} countermeasure by configuring the PRO to oscillate at randomized oscillation frequencies. Besides, the user can monitor the oscillation frequency of each PRO in the array by reading out its corresponding counter value. We demonstrate that by monitoring the frequency change of the PROs, on-chip local power attacks and EM fault injections can be detected. 

The proposed design can be used on any secure module, from small hardware accelerators to complex \acp{soc}. To the best of our knowledge, this is the first work to comprehensively study the potential of RO-based designs in SCA countermeasure, power sensing, and fault detection. 

\subsection*{Adversary Model}

PRO covers adversaries with side-channel and fault attack capabilities listed in the following:\\ 
\textbf{Side-channel Attacker Model.} We consider two attacker models. The first attacker model has physical access to the device which enables the attacker to control the input data and monitor the power dissipation by shunting the device's power supply. The second attacker model works remotely; The attacker circuit shares a \ac{pdn} with the victim circuit and can control \textit{only the attacker circuit} remotely. Therefore, the attacker is able to implement malicious logic to monitor the changes on the shared \ac{pdn} and measure the power consumption of the device \cite{zhao2018fpga,standaert2006overview}. This enables the attacker to perform side-channel attacks, such as Simple Power Analysis (SPA) \cite{mangard2008power}, Differential Power Analysis (DPA) \cite{Koc96}, and Correlation Power Analysis \cite{brier2004correlation}, to retrieve the secret information used in the victim circuit.\\ 
\textbf{Fault Attacker Model.} We also assume the adversary can induce faults into the victim circuit by stressing the electrical environment, such as injecting clock glitch, power glitch, and EM glitch. These glitches can induce targeted transient faults which can flip bits, change the control flow of the secure algorithm, set/reset the circuit, etc. Fault injection can be done either by exerting disturbance to the circuit directly, which requires the adversary to have physical access to the device, or by having remote access to the shared cloud computing environment with the victim circuit \cite{8715263,krautter2018fpgahammer,9237147, 8844478}.   The exact fault effects to the circuit highly depends on the fault injection parameters, victim circuit's architecture and algorithm, and fault injection technique. By monitoring the fault response of the circuit after injecting targeted faults, the adversary can retrieve the secret information by performing Differential Fault Analysis (DFA) \cite{biham1997differential}, Statistical Fault Analysis (SFA) \cite{fuhr2013fault}, or instruction skip attacks \cite{yao2018low}.
PRO as a secure on-chip add-on can be integrated to the circuit to protect against the aforementioned attackers. Adversaries may try to tamper with the PRO sensor itself to bypass the PRO's security mechanisms, but we don't consider this adversary model within this work.

The structure of the paper is as follows. The next section reviews related work of Ring Oscillators and highlights our contribution. \autoref{sec:PRO-design} describes our proposed PRO design. In \autoref{sec:SCA countermeasure}, we explain and demonstrate the effectiveness of PRO as a side-channel countermeasure. Next, we present the PRO's power sensing functionality in \autoref{sec:power sensing}. We further show that PRO can detect power fault and EM fault in \autoref{sec:fault-detection}. Finally, we conclude the paper in \autoref{sec:conclusion}.

\section{Related Work}\label{sec:RelatedWork}

When sharing the same \ac{pdn}, seemingly unsuspecting parts of the implemented logic can perform adversarial operations on the other parts.
In this work, our focus is on two categories of adversarial operations; fault injection and power side-channel analysis. 
In the following, we categorize the related work into three parts: using on-chip logic as a countermeasure against power \ac{sca}, using on-chip sensors as power sensors to detect power perturbation, and using on-chip sensors to detect fault injection attacks.
\subsection{On-chip sensors as a countermeasure against power \ac{sca}}
Liu et al. \cite{5477174} use an array of \acp{ro}, randomly switched on and off, to dynamically hide the power consumption of AES SBox and hinder the first-order \ac{dpa}.
Similarly, Krautter et al. \cite{8942094} use \acp{ro} as a power-based \ac{sca} mitigation methodology. In their work, the part of the implementation that needs to be protected is surrounded by a network of \acp{ro}. By switching an arbitrary number of the \acp{ro} on and off, the \ac{snr} in power traces decreases, and therefore, the number of traces required for a \ac{cpa} attack to be successful is increased. This approach is called {\em hiding} side-channel leakage. However, the RO in both designs are running at a fixed oscillation frequency, and thus, only a single-frequency noise is injected. In this case, it is straightforward for an attacker to apply post-processing techniques to remove the noise effect. To avoid this weakness, PRO uses user-controlled but random frequency changes (\autoref{sec:SCA countermeasure}). 
Moreover, to further reduce the overhead, we show how a simple modification can enhance the countermeasure efficacy. 
\subsection{On-chip sensors to detect/cause power perturbation}
 

Zick et al. \cite{zick2012low} use \ac{ro}s to measure on-chip voltage variations. Indeed, the oscillation frequency is proportional to the supplied voltage on the \ac{pdn}. To measure the frequency of an \ac{ro} accurately, counters are required that are clocked with the output of the \ac{ro}. This limits the maximum sample rate attainable by the \ac{ro} counter structure, and hence, the bandwidth of the side-channel signal. 
This limitation has motivated research on other voltage-sensitive \ac{tdc} methods.
For instance, Gnad et al. \cite{7929182} use carry-chain primitives available on Xilinx FPGAs as \ac{tdc}s. However, the use of carry-chain primitives makes their approach specific to certain FPGA families.
Similar \ac{tdc} structures have been explored in the context of CMOS design simulation to measure the operating voltage of a chip \cite{anik2020chip}.
 
Moreover, \ac{ro}s have been used in offensive scenarios affecting the \ac{pdn} for both passive (power-based) and active (fault injection-based) physical attacks.
As an example of power-based \ac{sca}, Zhao et al. \cite{zhao2018fpga} presented on-chip power monitors with \ac{ro}s. They demonstrated that \ac{ro}s can be used as a power monitor to observe the power consumption of other modules on the FPGA or SoC. Using their power monitor, they captured power traces of the device running the RSA algorithm and were able to successfully find the private key by applying \ac{spa}. 
Gravellier et al. \cite{8994789} perform \ac{cpa} on power traces acquired with \ac{ro}-based power sensors.
 
 To inject timing faults, Mahmoud et al. \cite{8715263} employ \acp{ro} to increase the voltage drop on the power network and lower the voltage level. Effectively, they make the victim chip slower, causing timing faults. Similar attacks have been shown in other works \cite{krautter2018fpgahammer,9237147, 8844478}.

\subsection{On-chip sensors to detect fault injection}

Next, we consider on-chip sensors for fault detection.
Miura et al. \cite{7544333} present a sensor consisting of \ac{pll} and \acp{ro}. In their work, \acp{ro} are routed in a specific way to ensure their path travels through most parts of the chip. Once an EM fault is injected, the path delay of the \acp{ro} will be affected, resulting in changes in the \ac{ro} phase. The \ac{pll} logic can capture this phase disturbance and detect the ongoing fault injection. Similarly, He et al. used PLL block to detect the laser disturbance on RO oscillation frequency \cite{he2016ring}. 

Provelengios et al. \cite{9237147} show that on-chip \ac{ro}s can not only detect fault injection, but also locate the origin of the fault injection. With a similar structure, RON \cite{zhang2011ron} builds a ring oscillator network, distributed across the entire chip, to detect hardware Trojans. 
 Their work confirmed that \ac{ro}-based power sensors can have a sufficiently high sample rate to detect fluctuations on the \ac{pdn}.

 However, the scope of their work is limited to the power fault detection, 
 whereas, in our work, we further investigate EM fault detection (\autoref{sec:fault-detection}). Additionally, the unique programmable design of our proposed RO structure also enables its usage for power SCA countermeasure (\autoref{sec:SCA countermeasure}).



\subsection{Our contribution}
In general, each previous work addresses one single aspect at a time: a side-channel countermeasure, a power monitor, or a fault detector. In practice, an adversary is capable of performing a combination of attacks. Hence, it is crucial to find a security mechanism that encapsulates protection against these attacks. 
Our goal in this work is therefore to design a \textit{programmable} \ac{ro} structure that can provide the following functionalities within the same structure:
\begin{enumerate}
    \item Hiding protection against power-based \ac{sca}.
    \item On-chip power monitoring of the fluctuations on the \ac{pdn}.
    \item Detecting fault injection. 
\end{enumerate}
To the best of our knowledge, this is the first work to comprehensively investigate the RO's potential in addressing all these three aspects. 
In the following sections, we introduce our proposed design and demonstrate through experiments the capability of the proposed system.
Even though we demonstrate our experiments as an FPGA prototype, our design is not limited to FPGAs and can be extended to other electronic chips.  

\section{Programmable RO Design} \label{sec:PRO-design}
\subsection{Background}
\begin{figure}
    \centering
    \includegraphics[width=.6\linewidth]{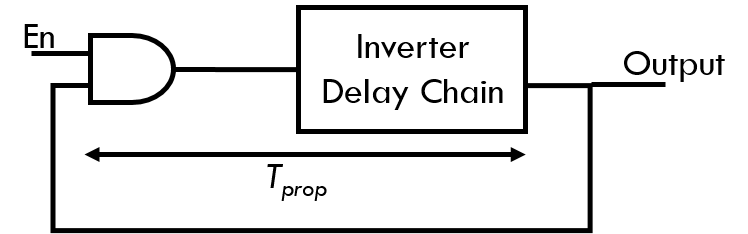}
    \caption{Propagation delay of a ring oscillator.}
    \label{fig:t_prop}
\end{figure}
In this section, we introduce our Programmable RO (PRO) sensor design. As shown in the \autoref{fig:t_prop}, Ring Oscillator (RO)'s output oscillation frequency depends on the propagation delay of their internal signals. In each oscillation period of a RO, the signal has to propagate twice through the propagation path. 
Therefore, the oscillation period ($T_{RO}$) of a RO is $T_{RO} = 2 \cdot T_{prop}$ and its frequency follows the following equation:
\begin{equation} \label{ro_freq}
    f_{RO} = \frac{1}{2 \cdot T_{prop}}
\end{equation}
More specifically, the propagation delay path is composed of an odd number of inverters and each inverter contributes to the delay of the path. If \(t\) represents the delay of an individual inverter, and \(n\) denotes the number of inverters in the chain, its frequency follows the following equation:
\begin{equation} \label{ro_freq_2}
    f_{RO} = \frac{1}{2n\cdot t}
\end{equation}
Hence, the frequency of a RO can be controlled by adjusting the number of stages in the inverter chain.
\subsection{PRO Design and Configuration}

In this work, we aim to have a programmable design of the RO which gives the designer the flexibility to choose the RO oscillation frequency.

\begin{figure}[t]
    \centering
    \includegraphics[width=.9\linewidth]{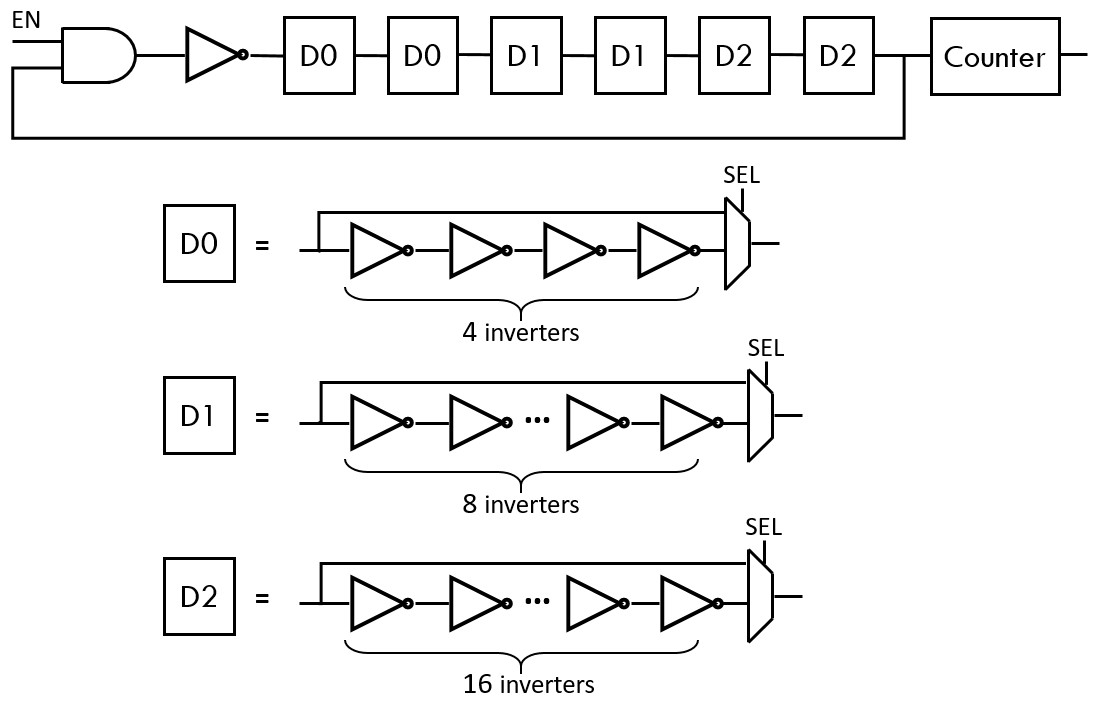}
    \caption{PRO Design. D0 donates the delay cell type-0, D1 donates the delay cell type-1, D2 donates the delay cell type-2.}
    \label{fig:ro_structure}
\end{figure}

\autoref{fig:ro_structure} shows the basic structure of our proposed design of the programmable sensor. The PRO consists of multiple delay cells. Each delay cell includes two delay paths; one consisting of inverters and the other a shorting path which bypasses the inverters. The multiplexer in the delay cell can control the delay cell's propagation delay by selecting between the delay path and the shorting path with the control input signal \texttt{SEL}. Each delay cell has its independent control signal. Suppose there are \(N\) inverters configured in the delay cell, when \texttt{SEL} = 1, the delay path is selected and when \texttt{SEL} = 0, the shorting path is selected. The propagation delay of each delay cell \(T_{C}\) is therefore: 
\begin{equation} \label{delaycell_freq}
    T_{C} = SEL\cdot T_{d} + (1-SEL)\cdot T_{s}  
\end{equation}
Where \(T_{d}\) denotes the propagation delay of the delay path and  \(T_{s}\) denotes propagation delay of the shorting path. The propagation delay of the shorting path \(T_{s}\) is a very small value compared to \(T_{d}\) but not 0, this is because of the delay of routing and the delay of the multiplexer. Other user control inputs include \texttt{EN} which controls whether PRO is enabled (oscillating) or not, and a control signal to reset/read the PRO counter. The structure of the PRO design gives flexibility to the designer in manifold. As shown in \autoref{table-PRO-config}, there are multiple initial structural configurations to be decided by the hardware designer at design time, including the number of inverters per delay cell, as well as the number and type of different delay cells.
These parameters determine the range of the programmable RO's oscillation frequency and the number of frequency configurations the programmable RO can have.

Several constraints can be used as the guidance while configuring the Initial Design Configurations of PRO: 
\begin{enumerate}
    \item Oscillation frequency range;
    \item Number of configurations;
    \item Size of frequency changing step;
    \item Area;
\end{enumerate}
As a starting point of PRO parameter configuration, the designer should estimate the propagation delay $T_{prop}$ for a single inverter. This knowledge can be obtained through the design library, timing simulation, or measuring RO's oscillation frequency with a single inverter (when working on an FPGA environment). Then, based on the designated Oscillation frequency range of PRO, the designer can calculate the minimum and maximum number of inverters are needed by \autoref{ro_freq_2}. After deciding the number of inverters needed, the designer can group the inverters into different types of delay cells based on the designated frequency changing step and number of configurations that are needed. Theoretically, more inverters result in a larger oscillation frequency range at a cost of larger PRO area. Therefore, based on the targeted protect design area, the designer should decide the area constraint for the PRO, so that the each PRO can have a good spatial coverage of the design while at the same time wouldn't be too close to influence other PROs' local power distribution.


\begin{table}[]
\centering
\caption{Configurations for PRO}
\label{table-PRO-config}
\begin{tabular}{lllll}
\cline{1-2}
\multicolumn{1}{|l|}{\textbf{Configuration Type}}            & \multicolumn{1}{l|}{\textbf{Configurations}}                                                                                                                 &  &  &  \\ \cline{1-2}
\multicolumn{1}{|l|}{Initial Design Configurations} & \multicolumn{1}{l|}{\begin{tabular}[c]{@{}l@{}}number of delay cell type, \\ number of delay cells,\\ number of stages in delay cells\end{tabular}} &  &  &  \\ \cline{1-2}
\multicolumn{1}{|l|}{User Configurations}           & \multicolumn{1}{l|}{\begin{tabular}[c]{@{}l@{}}\texttt{EN}, \\\texttt{SELs}, \\ Counter Start/Stop\end{tabular}}                                                     &  &  &  \\ \cline{1-2}
                                                    &                                                                                                                                                     &  &  & 
\end{tabular}
\end{table}

Next, to better explain our proposed structure, we pick one configuration as an example. \autoref{fig:ro_structure} shows the structure of the  PRO with three types of delay cells. The type-0 delay cell (D0) has 4 inverters, the type-1 delay cell (D0) has 8 inverters and type-2 delay cell (D0) has 16 inverters. We instantiated 2 of each type of delay cells in the inverter chain. All the delay cells have an even number of inverters, and 1 inverter is instantiated at the start of the inverter chain to make sure that there is always an odd number of inverters in the inverter chain.  When all the inverters are configured to be used in the delay path, the propagation delay \(T_{prop}\) is maximal, therefore, the overall programmable RO will oscillate at its lowest frequency. When all the delay cells are configured to use the shorting path, the propagation delay \(T_{prop}\) is minimal, therefore, the overall programmable RO will oscillate at its highest frequency. 

In our experiment setup, we implement PRO on Xilinx Spartan-6 FPGA, which is fabricated with 45nm CMOS technology. Under the aforementioned configuration, we measured that the lowest oscillation frequency is 22MHz and the highest oscillation frequency is 123.44MHz.  
Since each delay cell's {\tt SEL} is independent, there are in total 15 frequency configurations consisting of \{1, 5, 9, ..., 57\} inverters.
Since there are six {\tt SEL} signals, there are 64 configurations in total which redundantly map into the 15 achievable configurations. Through this redundancy, we are able to estimate the local manufacturing process variations, which is helpful to decide when a deviation should be cause for alarm (i.e., fault detection) or not.


The designers can control the RO's frequency by setting the input value of {\tt SEL}. We are using the same configurations for all the later experiments in this paper. For under this PRO configuration, each PRO can be implemented with 128 LUTs and 32 Registers, in total 160 slices. 
In our experimental setup, a PRO array with 36 PROs can cover the whole FPGA (46648 LUTS and 93296 Registers, in total 139944 slices ) to provide the whole chip power monitoring and fault detection. Therefore, only an overhead of 4.1\% is introduced.




\begin{figure}[t]
    \centering
    \includegraphics[width=.8\linewidth]{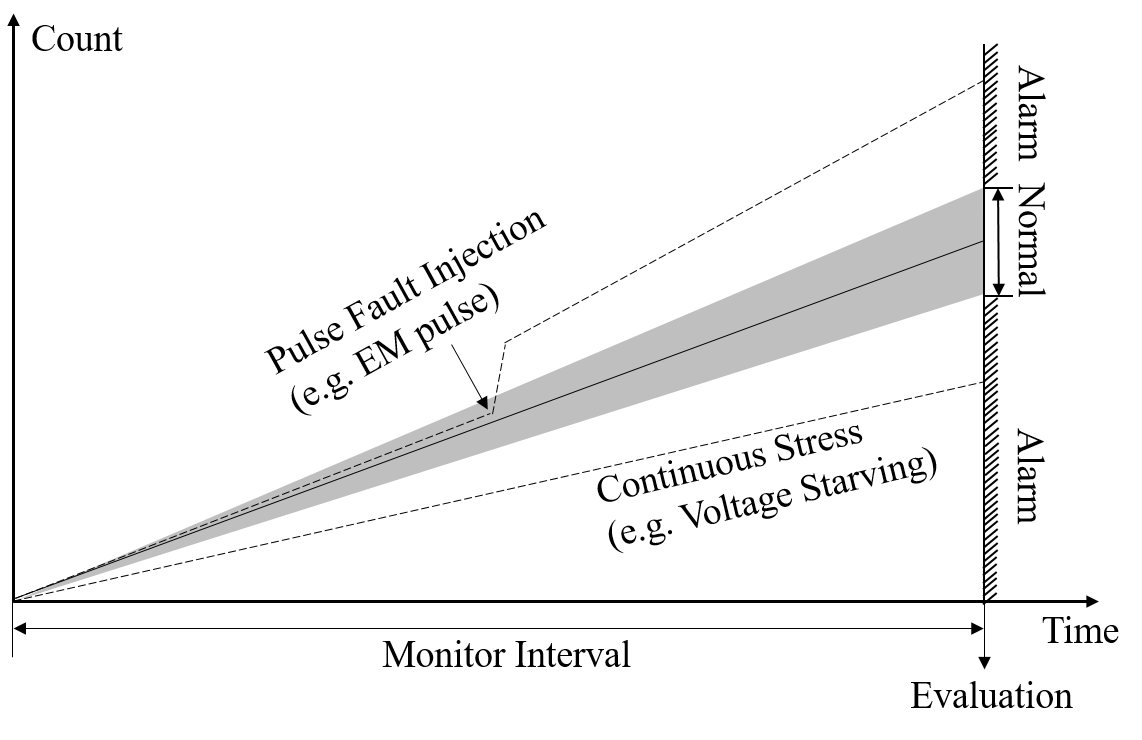}
    \caption{Basic principles for PRO fault detection}
    \label{fig:PRO_basic_principle}
\end{figure}


\subsection{PRO Integration and Basic Principles}

As a security resistance add-on, PRO can be integrated into the design to protect simple designs such as hardware encryption engines as well as complex systems such as an \ac{soc}. Control signals are needed for communicating with the PRO. The control signals set up the user control configurations in \autoref{table-PRO-config}. Generally, different control mechanisms can be adopted by the designer. In an SoC, the designer can add PROs as a co-processor which can be controlled by the processor through memory-mapped registers. Under this environment, the software running on the processor can configure the PROs on the chip. Therefore, PRO-based countermeasures can be dynamically enabled/disabled while the software is running. Besides, a hardware-based Finite State Machine (FSM) can be used to control the PROs as well. In our experiments, we are using the UART protocol to communicate with PRO, and the control signals sent through the UART are generated by a python script in this paper. 

\autoref{fig:PRO_basic_principle} shows the high-level basic principles for the fault detection mechanism using the PROs. The counter value will be evaluated at the end of each monitoring interval and compared with the reference counter value to get the actual oscillation frequency of the PRO. Under normal circumstances, each PRO oscillates at a certain constant frequency, and thus, its counter value will increase linearly during the monitoring interval. There will be some small variances caused by the environmental changes, jitters, process variance of the manufacturer, etc. A characterization procedure, therefore, is needed to define the range of normal operation \cite{tajik2017pufmon}. However, in the occurrence of instant fault injection (e.g. power glitch, EM pulse, time glitch, laser pulse), the counter will be disturbed. The counter value read out at the evaluation time will deviate from the normal range, and thus, a pulse fault injection will be detected by the PROs. Additionally, an adversary can inject timing faults by stressing the PDN continuously (e.g.power starving). As a result, the victim circuit will operate slower and cause timing violations to create faults. In this case, the PRO counter value will also deviate from the normal value and capture the fault injection event. In this paper, we use power fault and EM fault as an illustration, but PRO's fault detection coverage is not only limited to these two fault types.  


\section{Side-channel Countermeasure}\label{sec:SCA countermeasure}

Masking and hiding are two popular techniques for side-channel countermeasures. In masking, each secret variable is split into two or more shares which are concealed by random numbers. The side-channel leakage of each share alone does not reveal the secret variable because of the randomization introduced by random numbers. A random source that provides fresh random variables is significantly important in masking implementations.  Hiding countermeasures reduce the \ac{snr} for secret data-dependent operations. Hiding can be achieved by several techniques, such as by reshuffling cryptographic operations in a data-dependency consistent but random order \cite{tillich2008attacking}, inserting random delays \cite{coron2010analysis}, and running multiple tasks in parallel \cite{standaert2006overview}. In this work, we utilize the proposed PRO design as a hiding countermeasure by injecting noise with random frequency.  Previous work has proposed injecting noise for reducing the \ac{snr} \cite{das2017high} \cite{liu2010low}. However, since only single-frequency noises are injected, it is not tricky for an attacker to decrease the effect of noise either by using a band-pass filter while collecting traces or by post-processing the collected power traces, such as applying averaging, filtering, and frequency domain analysis. Thus previously proposed noise-injection-based hiding mechanisms still have security flaws. In our proposed design, we inject random-frequency noises with the PRO design so that it will be much 
harder for an adversary to eliminate the noise. 

\begin{figure}[h]
  \centering
  \includegraphics[width=\columnwidth]{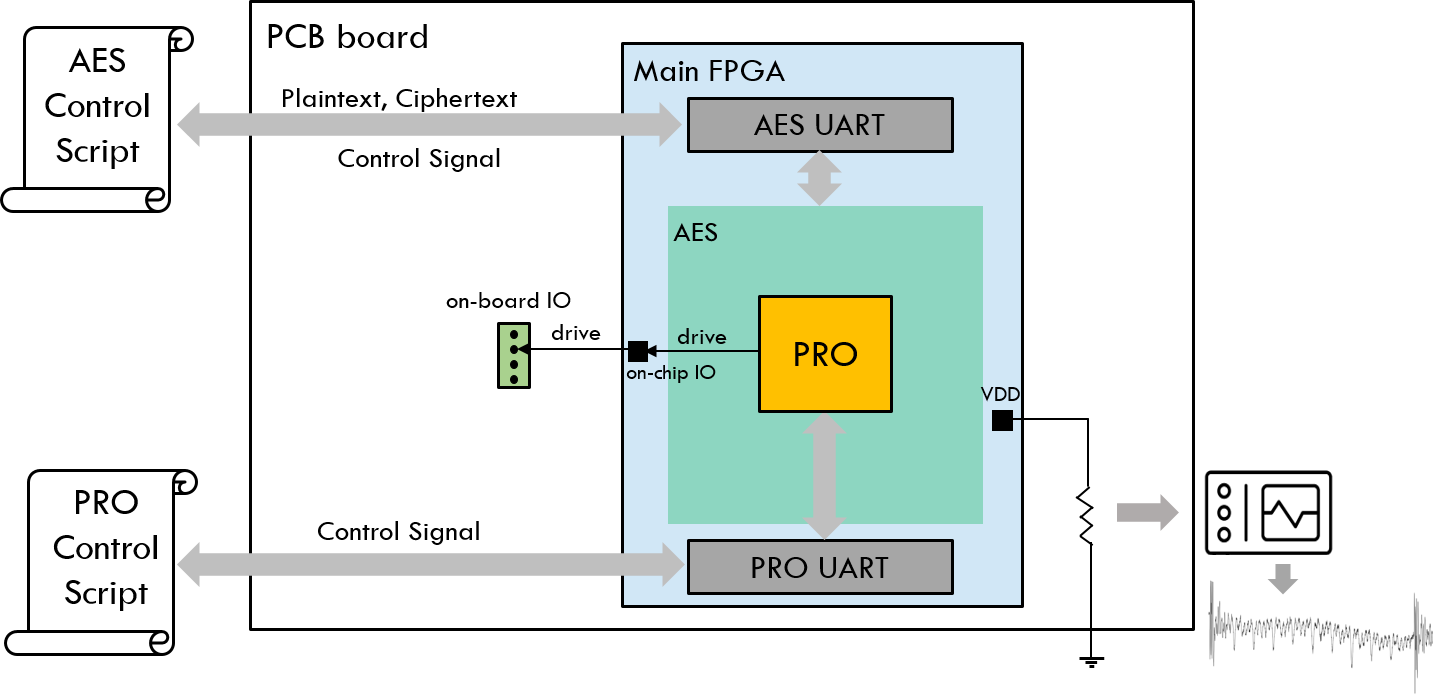}
  \caption{Experimental Setup for Evaluating RO's performance in side-channel leakage hiding}
  \label{fig:power_setup}
\end{figure}

The countermeasure circuit consists of a single PRO whose frequency can be controlled by the \texttt{SEL} input signals. The PRO drives one of the IO pins on the board. As demonstrated in previous work \cite{liu2010low} \cite{krautter2020cpamap}, the power consumed by a single RO is not large enough to have a significant influence on the power profile of a complete chip or a complete cipher. Instead, hundreds of RO need to be instantiated on the chip to have a profound hiding influence. This approach will cause significant design overhead and has the potential risk of inducing power fault to the circuit \cite{kim2007faults}. In our proposed mechanism, by driving an I/O pin with a PRO, the effect of a single (randomly-switched) PRO to influence the off-chip power network is amplified. Since the load capacitance of an IO-pin is much larger than the load capacitance of an internal FPGA net, the IO-pin requires more energy to charge and discharge. In this manner, even with a single \ac{pro}, significant additional power is consumed to change the power consumption characteristic. In practice, an adversary senses the on-chip power consumption using a probe, either by connecting an external probe to the system via a power supply pin \cite{utyamishev2018real} or else using an EM probe. Both of these are dependent on the off-chip power network, and therefore, affecting the off-chip power network is an important factor to defeat an attacker maliciously monitoring the power profile. 


\begin{figure}[t]
  \centering
  \includegraphics[width=\columnwidth]{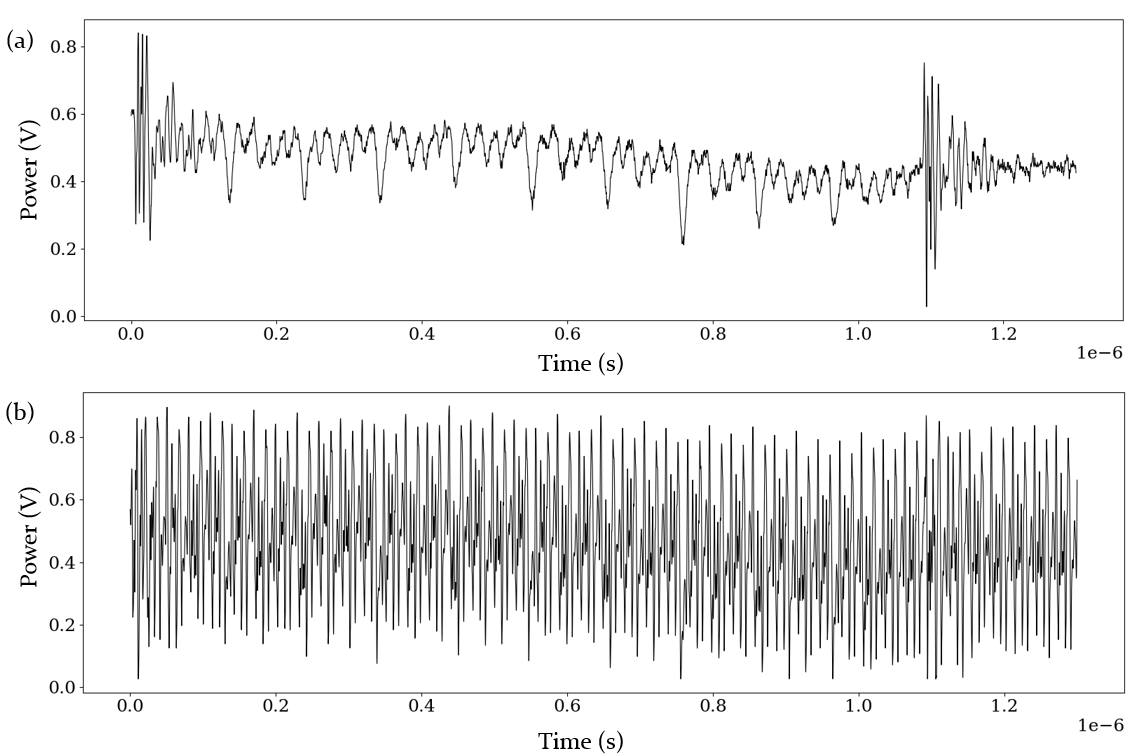}
  \caption{AES power traces when PRO is (a) Off; (b) On; }
  \label{fig:power_traces}
\end{figure}

\begin{figure}[h]
  \centering
  \includegraphics[width=\columnwidth]{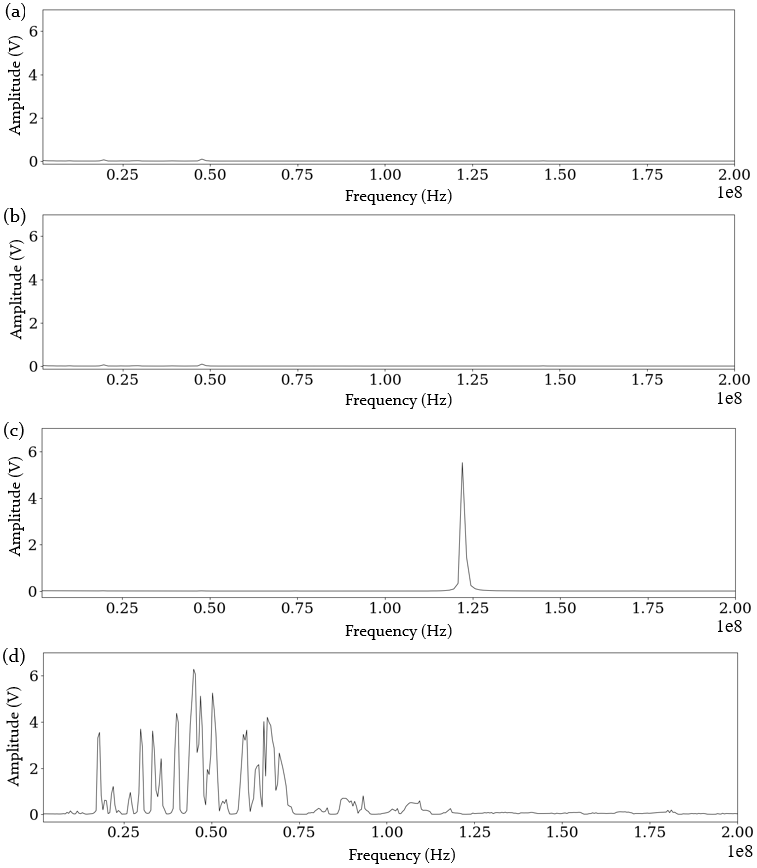}
  \caption{Power Spectrum for power traces when (a) PRO off; (b) PRO on without driving IO pin; (c) PRO with fixed oscillation frequency and driving IO pin; (d) PRO with random oscillation frequency and driving IO pin; 
  }
  \label{fig:power_spectrum}
\end{figure}

\begin{figure}[h]
  \centering
  \includegraphics[width=\columnwidth]{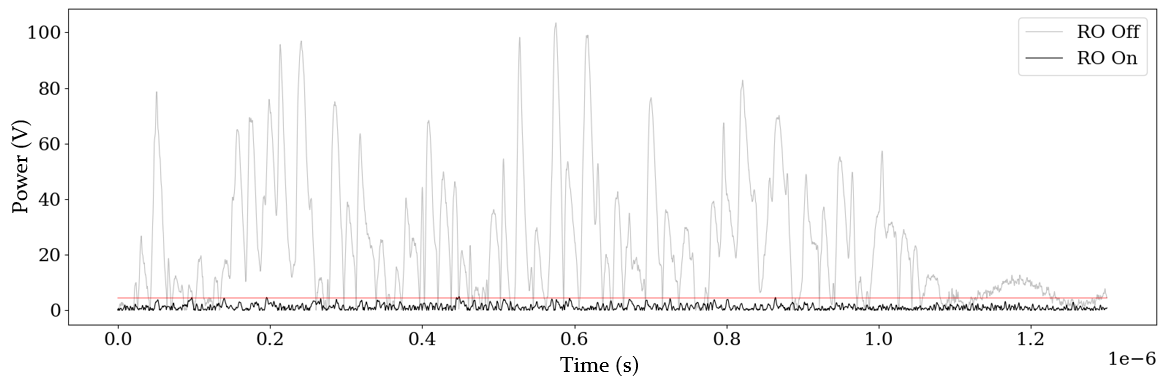}
  \caption{T-value Comparison when PRO is on/off }
  \label{fig:t_comp}
\end{figure}

The performance of our proposed hiding countermeasure design is evaluated with AES. \autoref{fig:power_setup} shows our experimental setup. We put hardware AES as well as the programmable sensor on the FPGA. The output signal of the PRO is mapped to drive the IO pin to amplify the noise effect. For each encryption scenario, plaintext and ciphertext are provided through the UART for AES. The communication procedure is controlled by the AES control script. At the same time, we use the sensor control script to send in control signals through the PRO UART. The control signals can enable/disable the RO and configure the oscillation frequency of the RO. While AES is running, the sensor's control script generates random numbers for the frequency configuration so that the frequency of the PRO can change randomly. Equally, an on-chip \ac{prng} can be used for this purpose.
\autoref{fig:power_traces}(a) shows the collected AES power trace when the programmable sensor is off. We can clearly see the pattern of ten rounds of the AES algorithm. By comparison, the power trace changes to a repeated oscillation pattern when we turned the PRO on, as shown in \autoref{fig:power_traces}(b), which indicates the strong influence of PRO on the power profile. Under our setup, the complete AES takes 41ms, and we configure the PRO control script such that the frequency of the PRO changes every 2ms, which means that the PRO's frequency will change at least 20 times while AES is running. \autoref{fig:power_spectrum}(a) shows the frequency spectrum of the power traces when the PRO is off. We can observe small peaks at the clock frequency (24MHz). We do not observe a significant influence on the power spectrum if we only put a single PRO without driving the IO pin.  \autoref{fig:power_spectrum}(c)(d) shows the power spectrum when the PRO is on and driving the output pin. By comparing to \autoref{fig:power_spectrum}(a), a significant influence on the frequency spectrum of the power profile can be observed while PRO is on. \autoref{fig:power_spectrum}(c) shows a sharp peak when we fix the PRO's oscillation frequency to 120MHz. 

Suppose one tries to protect the secure component by injecting noise with a regular RO with a single oscillation frequency. It is easy for the attacker to implement frequency spectrum analysis, find the injected noise frequency, and apply the corresponding filter to eliminate the influence of the injected protection noise. As a sharp comparison, \autoref{fig:power_spectrum}(d) shows that when random frequency noise is injected by PRO the frequency spectrum is expanded within the PRO's oscillation range from 22Mhz to 123.44Mhz. This makes it much harder for the adversary to filter out the noise by post-processing. To further evaluate the effectiveness of the proposed design on increasing side-channel resistance, we applied TVLA \cite{gilbert2011testing} on 50k collected traces; As shown in \autoref{fig:t_comp}, a dramatic decrease of t-value can be observed when the PRO is turned on compared to when the PRO is off. This indicates that the PRO design can significantly reduce the side-channel leakage of the circuit. 

\textbf{Note.} Generally, even though the adversary is aware of the noise signal, since the noise is injected by PRO at a random frequency which also changes at a fast pace, it is exceedingly hard to remove its effect by normal post-processing techniques; The adversary needs to monitor both the power consumption and the output of the PRO simultaneously with sufficient precision and should be able to remove the part of power consumption related to the output pad's oscillation using noise-cancellation techniques, which requires high-end devices. Additionally, to have sufficient information and perform a successful side-channel analysis from the obtained side-channel traces, the sampling rate for side-channel attacks
has to be at least 2$\times$ the clock frequency (according to the Nyquist theorem). We suggest that while choosing the initial design configurations in \autoref{table-PRO-config}, the designers should adjust the configurations such that the oscillation frequency range of the PRO covers at least 3$\times$ the clock frequency. Under our experimental setup, the clock frequency is 24MHz. Therefore we configured PRO's oscillation frequency to 22MHz - 123.44MHz, which covers about 5$\times$ the clock frequency. As a result, the adversary will need a higher-end device with a much higher sampling frequency (at least 10$\times$ the clock frequency) to successfully apply the same side-channel attack. Hence, PRO as a hiding countermeasure makes it much harder to attack the circuit by largely elevating the technique bar for the adversaries.

\section{Power Sensing} \label{sec:power sensing}
In this section, we demonstrate the on-die power monitoring functionality of the proposed PRO design. Power integrity is essential to guarantee the nominal function of the circuit. Therefore, monitoring of the fluctuations on the PDN is critical to detect abnormalities. We first explore the PRO's oscillation frequency with regard to the external power deviation. Then, we look into the PRO's performance in terms of local power sensing on the PDN of the chip.

Electric circuits use \acp{pdn} to deliver power to the transistors in the circuit via external voltage regulators. \acp{pdn} are still affected by sudden current consumption changes despite these voltage regulators. Thus, the sudden change in the switching activity induces transient voltage drops in the \ac{pdn}. \acp{pdn} can be modeled using RLC networks. The transient voltage drop seen by the \ac{pdn} can be defined as follows

\begin{equation}
    V_{drop} = IR + L \diff {i}{t}
\end{equation}
Here, the IR drop is due to the resistive components of the \ac{pdn} and is dependent on the steady-state current I. The other term, $L\diff{i}{t}$, influences voltage drop due to the inductive components of the \ac{pdn} and is dependent on the changes in the current over time.
Hence, as soon as there is a high current consumption/variations due to some switching activities of the logic circuit, the voltage drop will increase. 


The propagation delay of signals is affected by the on-chip voltage level; Higher voltage levels increase the switching speeds of transistors, whereas lower voltage levels decrease them. Since the voltage level affects the propagation delay of signals, the immediate frequency of a ring oscillator can indicate the level of the voltage on a chip. 
We take advantage of this property in our proposed PRO sensor to monitor the integrity of the on-chip power network.
\vspace*{-1cm}
\subsection{PRO Power Sensing with Regard to External Power Variations}
\begin{figure}[t]
  \centering
  \includegraphics[width=0.8\columnwidth]{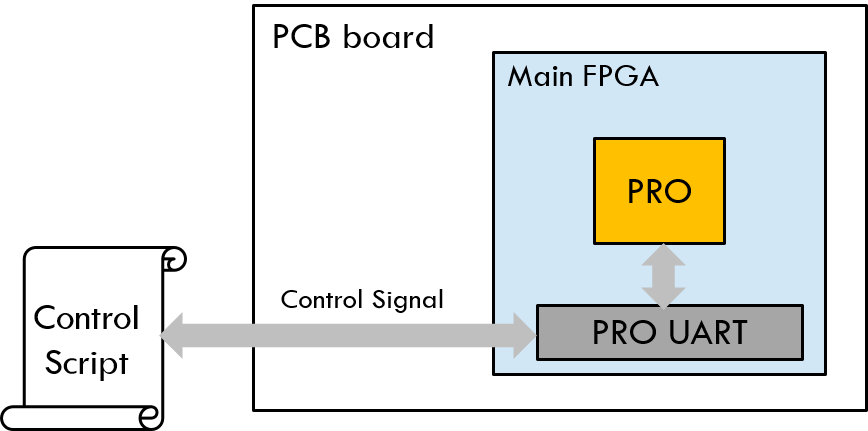}
  \caption{Experimental Setup for PRO frequency changing as a function of external power supply}
  \label{fig:ro_external_setup}
\end{figure}

\vspace*{-0.5cm}
We first investigate the PRO's frequency with respect to external power variations. \autoref{fig:ro_external_setup} shows the setup for this experiment scenario. We put a single PRO sensor on the FPGA.  For the PRO's frequency measurement, we start the PRO sensor's counter and system clock counter at the same time. After running for an arbitrary amount of time \(T_{arb}\), we read out the RO sensor's counter value \(C_{RO}\) and the reference system clock counter value \(C_{clk}\) through UART. Then, we calculate the PRO's frequency by:
\begin{equation} \label{freq_cal}
  f_{PRO} = \frac{C_{PRO}}{C_{clk}}\cdot f_{clk} 
\end{equation}
Where \(f_{PRO}\) is the PRO sensor oscillation frequency, \(f_{clk}\) is the reference clock frequency. We measure the value of $C_{PRO}$ 1000 times and take the average for better precision. The measurement procedure is automated through a control script running on a PC. 

As we mentioned in \autoref{sec:PRO-design}, the PRO's oscillation range is 22 Mhz to 123.44 Mhz. To investigate the PRO's power sensing sensitivity when operating under different frequencies, we set the PRO sensor to several oscillation frequencies at the starting (highest) power supply voltage for the main FPGA core (1.33V). The frequency configurations we pick are 153.2MHz, 100MHz, 66.8MHz, 40.5MHz, and 27.2MHz. We gradually decrease the FPGA's supply voltage and monitor the PRO sensor's oscillation frequency.  
\begin{figure}[H]
  \centering
  \includegraphics[width=0.9\columnwidth]{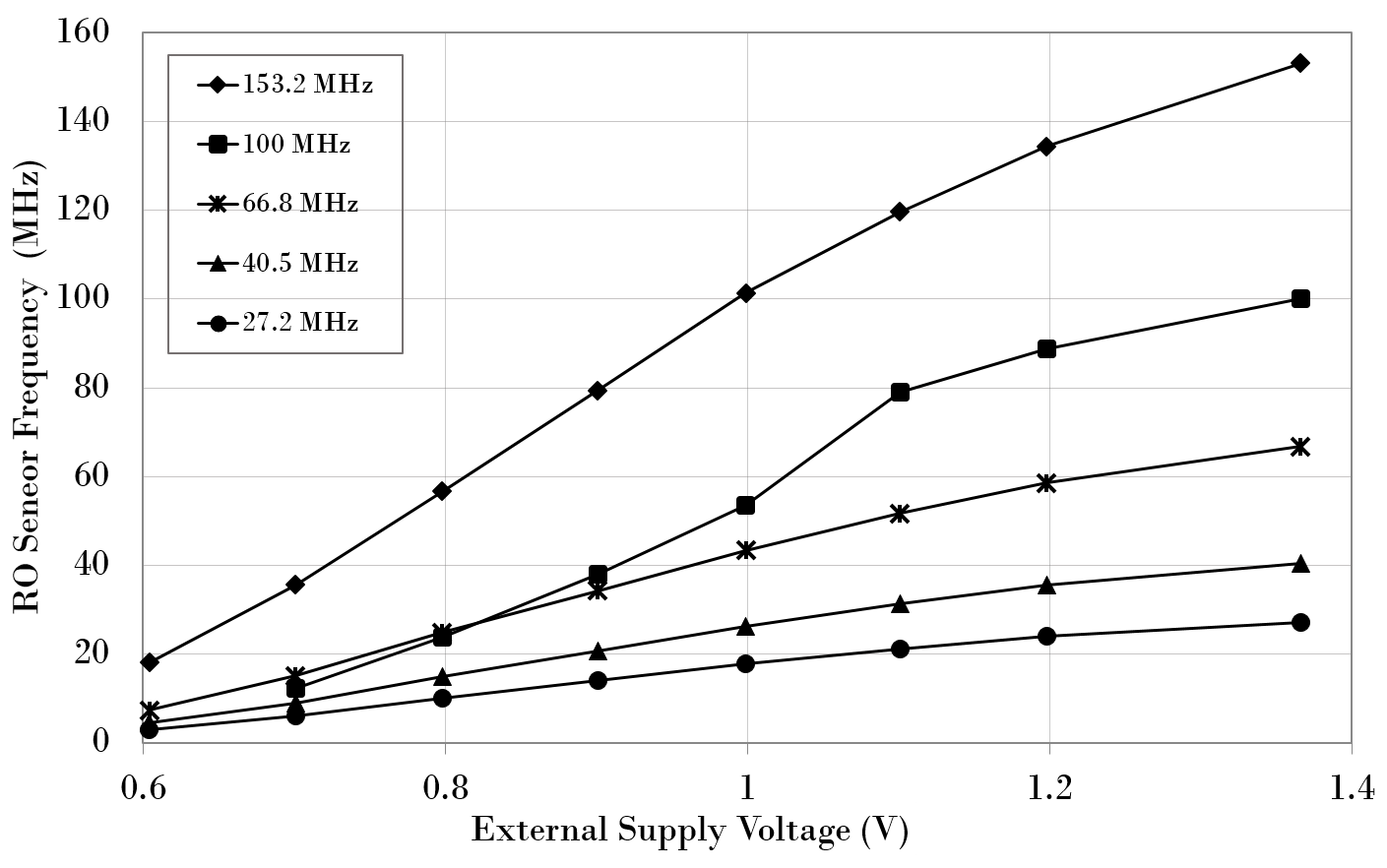}
  \caption{PRO's oscillation frequency with Regard of External Power Supply Voltage}
  \label{fig:ro_external}
\end{figure}

\autoref{fig:ro_external} shows the result of the PRO oscillation frequency with regard to the external supply voltage. As shown in the figure, when the external supply voltage drops, the PRO's frequency drops steadily. The PRO's oscillation frequency reflects the power supply voltage, and therefore, it can sense the changes of the power supply and can be used for power monitoring. With respect to the sensitivity of power sensing, it can be observed that the higher the oscillation frequency is, the sharper the slope of the frequency vs. the external supply voltage line will be. This indicates that a higher oscillation frequency can achieve higher sensitivity in detecting power variations.  

\subsection{PRO Power Sensing with Regard to On-die Local Power Variations}
\begin{figure}[h]
  \centering
  \includegraphics[width=0.6\columnwidth]{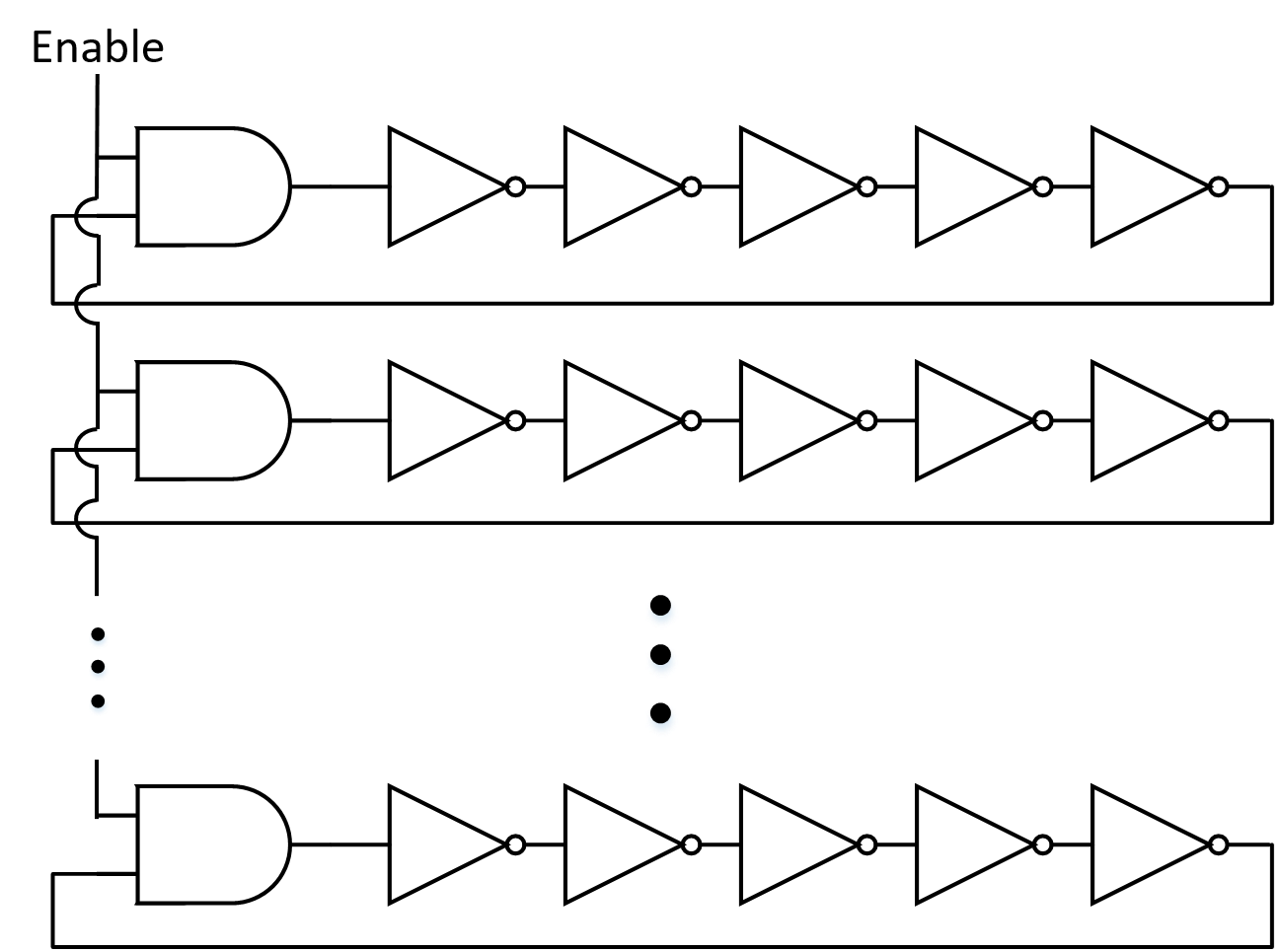}
  \caption{The structure of the employed RO-based power wasters.}
  \label{fig:power_wasters}
\end{figure}

After investigating the correspondence between the PRO sensor's oscillation frequency and the variations of external power variations, we evaluate the power sensing performance with regard to the on-die local power changing. Several previous works have shown that RO-based power wasters can cause a local power supply drop \cite{moini2020understanding} \cite{provelengios2019characterizing} \cite{zhao2018fpga}. This will cause the local circuit's logic to operate at a lower voltage, therefore the local power sensor should show a decrease in the oscillation frequency when the power wasters are turned on. In this work, we adopt the RO-based power waster shown in \autoref{fig:power_wasters}. 
Each power waster has five inverters in the delay chain with an \texttt{AND} gate and oscillates at 245MHz. 
A global enable signal is used to turn on/off all the power wasters in the circuit. 

\begin{figure}[h]
  \centering
  \includegraphics[width=0.9\columnwidth]{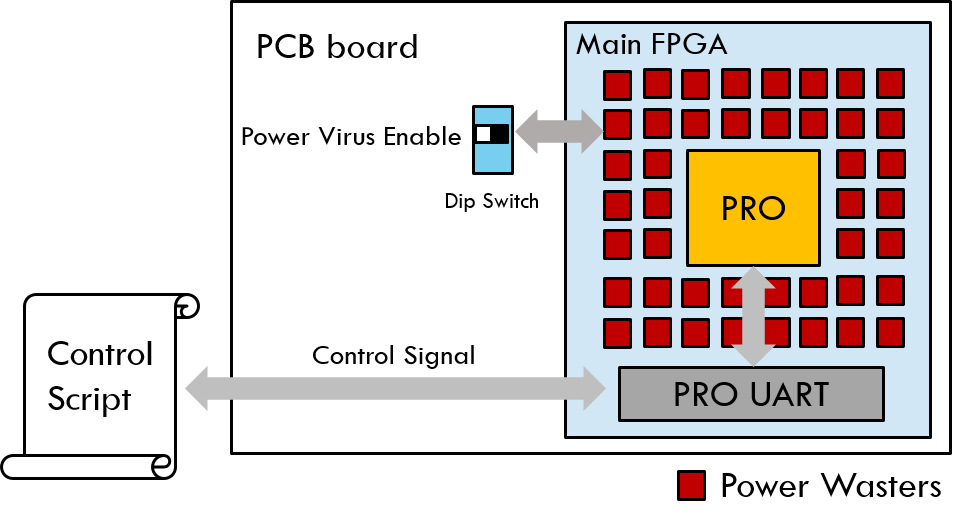}
  \caption{Experimental Setup for PRO Power Sensing with Regard to On-die Local Power Variations            }
  \label{fig:ro_local_setup}
\end{figure}

\begin{figure}[h]
  \centering
  \includegraphics[width=0.85\columnwidth]{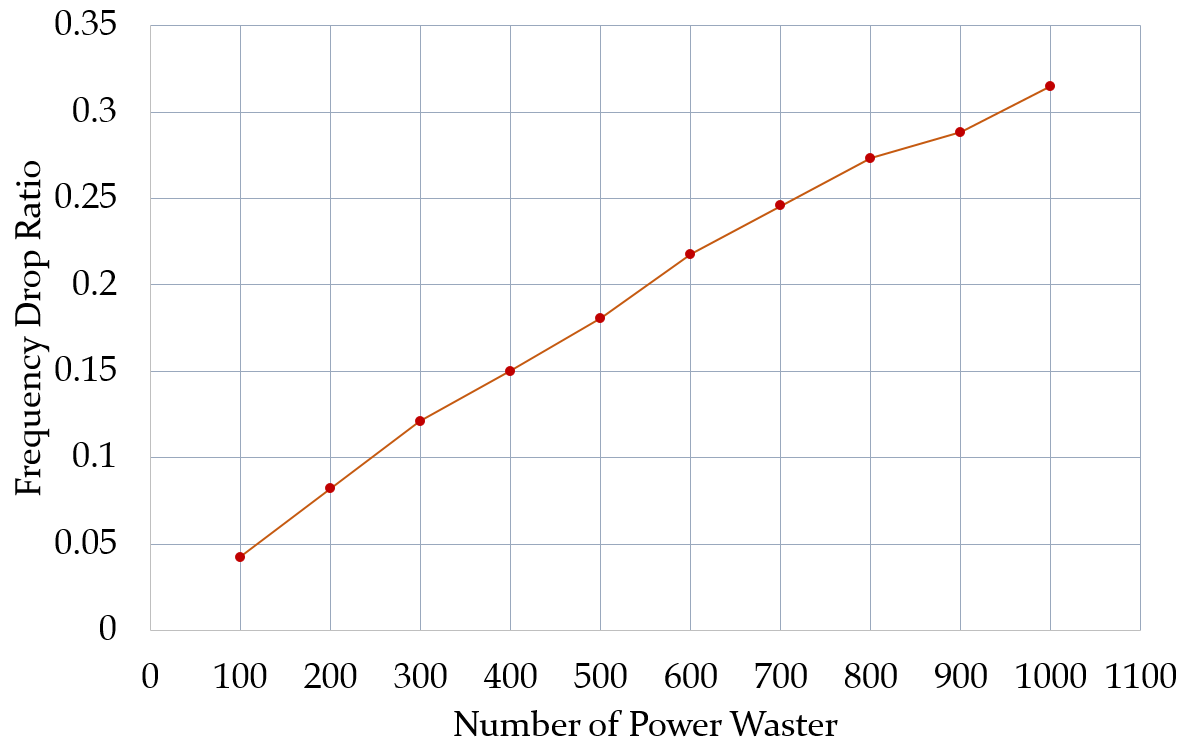}
  \caption{ PRO Frequency with Regard of Local Power Supply}
  \label{fig:ro_internal}
\end{figure}

\autoref{fig:ro_local_setup} shows the experimental setup for the local power sensing evaluation. In this setup, UART communication is used to read out PRO's counter value. 
We constrain the power waster to locate around the PRO sensor to induce the local power drop around the sensor. By configuring the number of power wasters, we can control the amount of local power drop. An on-board dip switch is used to enable/disable the power wasters. In a measurement scenario, we gradually increase the number of power wasters. For each number of power waster configuration, we measure the PRO's oscillation frequency 1000 times and take the average with power waster on/off, respectively. Next, we calculate the Frequency Drop Ratio as follows: 
\begin{equation} \label{freq_drop}
  Frequency\:Drop\:Ratio = \frac{f_{off} - f_{on}}{f_{off}} 
\end{equation}
In \autoref{freq_drop}, \(f_{off}\) denotes the PRO sensor's frequency when the power wasters are disabled (turned off) and \(f_{on}\) denotes its frequency when the power wasters are enabled (turned on). The results from the experiment are shown in \autoref{fig:ro_internal} when different numbers of power wasters are enabled. As more power wasters are enabled, the frequency drop ratio increases correspondingly. We can observe a nearly linear relationship between the number of power wasters and sensor oscillation slowdown. The linear regression which can closely model the correlation between the number of power wasters and the frequency drop ratio can be constructed as \(f(x) = 0.00031x + 0.247\)  with an R-squared value of 0.991. Therefore, we conclude that PRO can effectively sensing the local power variations as well.

\subsection{PRO Power Sensing with Regard to Sensor Locality}
\begin{figure}[h]
  \centering
  \includegraphics[width=0.6\columnwidth]{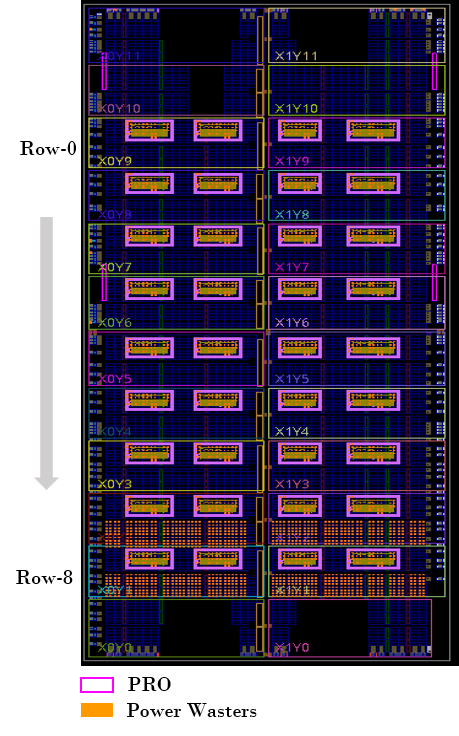}
  \caption{FPGA Floorplan for Evaluating PRO Performance with Regard of Sensor Locality}
  \label{fig:ro_locality_floorplan}
\end{figure}

In this experimental scenario, we evaluate the PRO sensor's frequency change with respect to the spatial proximity to the switch logic that consumes the power. In this experiment, we instantiate 36 PRO sensors to get full spatial coverage of the FPGA. As shown in the floorplan in \autoref{fig:ro_locality_floorplan} for this experimental scenario, 36 sensors residing in 9 rows, and each row has 4 sensors.

\begin{figure}[h]
  \centering
  \includegraphics[width=0.85\columnwidth]{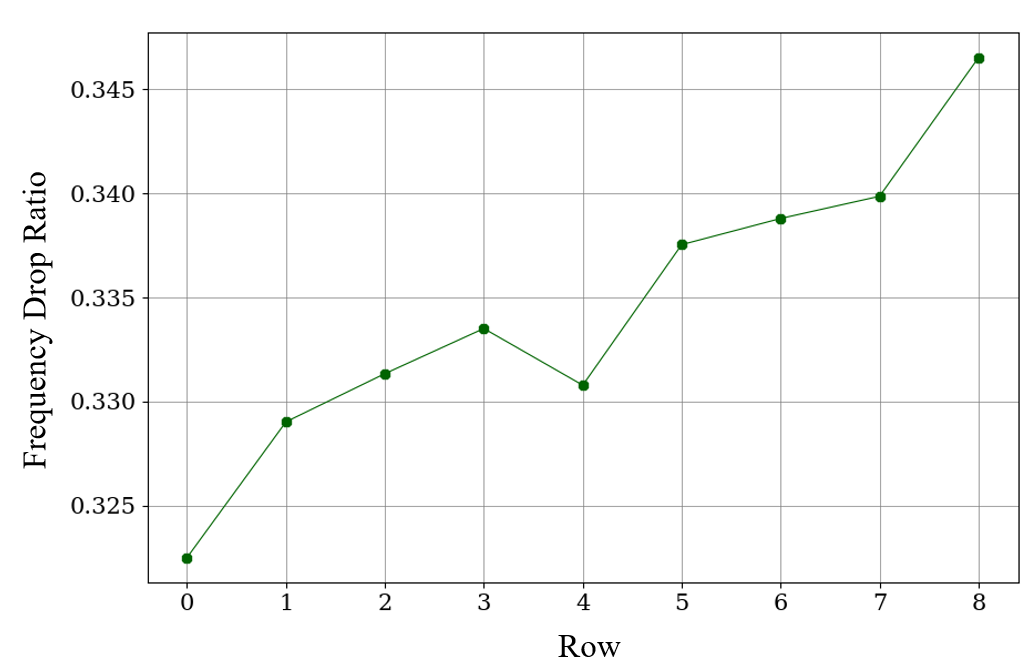}
  \caption{PRO's average Frequency Drop Ratio for each row versus the spatial proximity of the power wasters}
  \label{fig:ro_locality_Result}
\end{figure}

To remove the process variations among the PRO instances, we calculate the Frequency Drop Ratio for each PRO instance following \autoref{freq_drop}. We first measure the Frequency Drop Ratio for all the sensors. Then, we take the average of the frequency drop of the 4 PROs in each row.
The results are shown in \autoref{fig:ro_locality_Result}. We observe that as the PRO sensors are placed closer to the power wasters (from Row 0 to Row 8), the Frequency Drop Ratio increases. Therefore, we can see the spatial distance of the PRO sensor to the switching logic (power wasters) indeed can be reflected in the Frequency Drop Ratio. We can further use this feature to detect the location of injected faults on the chip (will be demonstrated in \autoref{sec:fault-detection}). Note that there is an outlier in our designed sensor, which might be attributed to the power distribution network structure of the electronic circuits in which the power in the center of the chip is built to be more stable \cite{popovich2007power}. 

\section{Fault Detection} \label{sec:fault-detection}
In this section, we focus on evaluating our proposed on-die PRO sensor's performance in sensing the occurrence of fault injection attacks. We show that PRO can be used to protect the circuit from adversaries who have physical access or remote control of the device \cite{8715263} which enables them to inject power or EM faults. However, we assume that PRO itself is protected against the manipulation of the attacker. We demonstrate that PRO can not only detect the occurrence of a power-based fault, but also the sensor array can detect the location of the power fault. This enables the designer (or the system administrator) to identify the source of the fault injection or the malicious circuits and build highly targeted fault response mechanisms accordingly. Moreover, We further demonstrate that PRO can be used for EM fault detection as well.

\subsection{Power Fault Detection}

\begin{figure}[h]
  \centering
  \includegraphics[width=0.9\columnwidth]{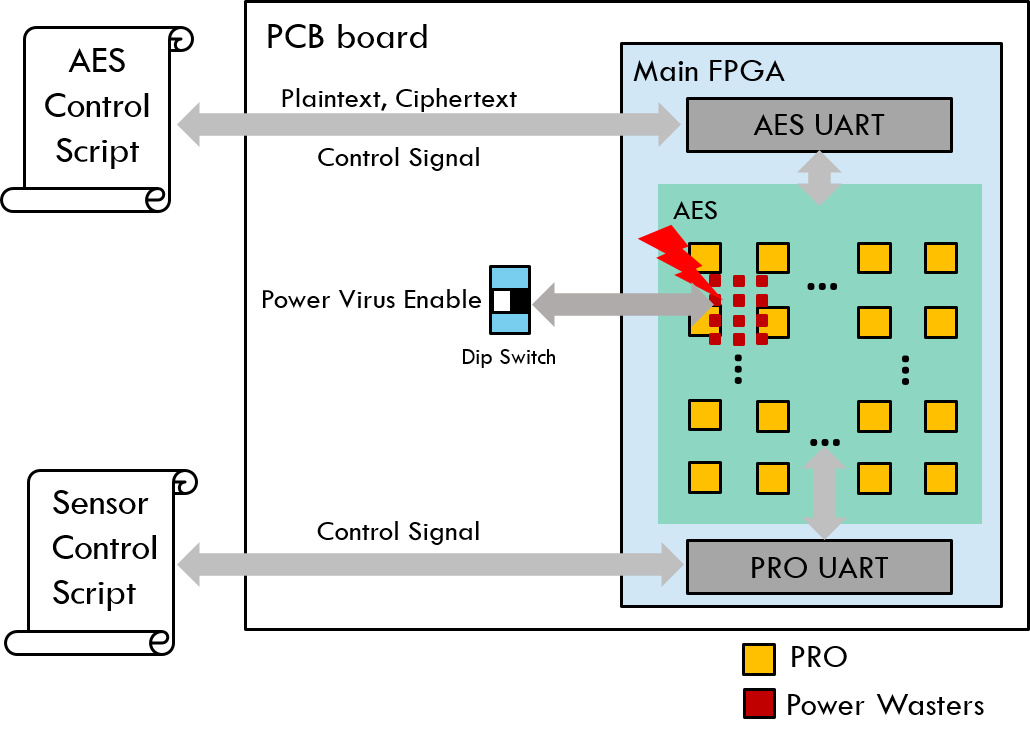}
  \caption{Experimental Setup for PRO Power Fault Detection}
  \label{fig:powerfault_setup}
\end{figure}

Sharing the same \ac{pdn} between a potential adversary and a victim opens the door to a new array of attacks. An adversarial logic can impose strong changes on the voltage level to cause timing faults in the victim circuit \cite{krautter2018fpgahammer,9237147, 8844478, 8715263}. Since all these attacks affect the \ac{pdn}, we aim to build sensors that are sufficiently sensitive to the voltage level and therefore can detect such attacks. Detecting ongoing fault injection attacks will prevent resulting timing faults to go unnoticed.

\autoref{fig:powerfault_setup} shows our experimental setup for evaluating the power fault detection performance of our sensor. We instantiate AES as well as the PRO sensors array on the FPGA. Power wasters are placed locally on the chip to simulate the situation when local power faults are induced by an adversary.  An on-board dip switch can control the turning on/off of the power wasters. AES control script is used to control starting the AES, send in plaintext, and read out the ciphertext. The AES control script is also used to monitor the correctness of the resulting ciphertext. We adjust the number of power wasters instantiated while AES is running. When faulty ciphertexts are observed, we know that an effective power fault is successfully injected. This ensures that the power fault detected by PRO are actually effective faults.  Next, we read out the PRO's counter value through the sensor's control script both when the fault is injected and not injected respectively, and compare their values. Note that as a chip-level sensor, our goal is to detect the location of the attacker instead of identifying the fault effect within the victim algorithm/circuit. 

\begin{figure}[h]
  \centering
  \includegraphics[width=0.6\columnwidth]{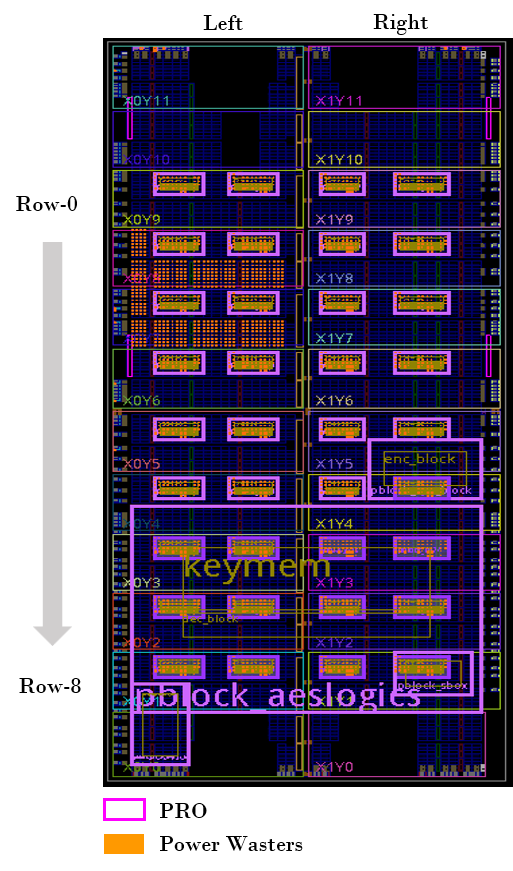}
  \caption{FPGA floorplan for Evaluating PRO Performance in Power Fault Detection, power wasters simulate local power fault happens at location-1.}
  \label{fig:powerfault_loc1}
\end{figure}

\begin{figure}[h]
  \centering
  \includegraphics[width=0.8\columnwidth]{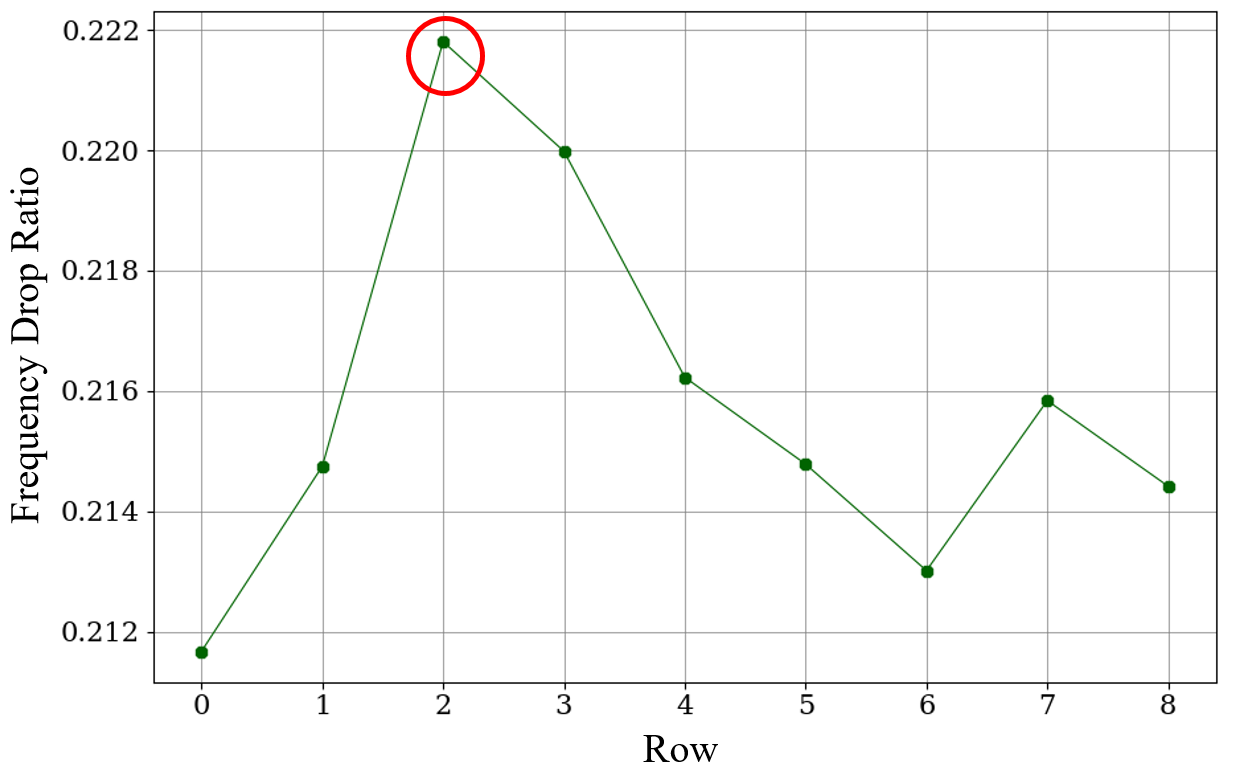}
  \caption{PRO average frequency drop ratio for each row when power fault happens at location-1}
  \label{fig:rowplot_loc1}
\end{figure}

\autoref{fig:powerfault_loc1} shows the floorplan of the aforementioned setup. We placed 36 sensors on the chip and 524 power wasters are instantiated to generate power fault. We first put the power wasters at Row 1 and Row 2 on the left as shown in the orange blocks in \autoref{fig:powerfault_loc1}. While AES is running, we read out the sensor's counter value when the faults are injected and not injected by power wasters respectively. Then we calculate the Frequency Drop Ratio based on \autoref{freq_drop}. With the PRO sensor data, we are to able find the location of the power fault. First, to locate which row has the power fault, we take the average of the 4 PRO sensors' frequency drop ratio in each row. \autoref{fig:rowplot_loc1} shows the result of each row's average Frequency Drop Ratio. The maximum Frequency Drop Ratio points to a location adjacent to Row 2. This demonstrates that our sensor array can point to the correct row that the fault has occurred. Then, we divide the chip into two regions, left and right. To locate the fault region, we take the average of the Frequency Drop Ratios of the 18 sensors in the left and right two columns separately. The average Frequency Drop Ratio on the left region is 0.2184, and the average frequency drop for the right region is 0.213. The left region is higher than the right region, which indicates that the source of the fault is in the left region. This demonstrates that our sensor array can point to the correct fault column. Now, after analyzing the data of the sensor array, we can locate the power fault's location at Row 2, left region.  

\begin{figure}[h]
  \centering
  \includegraphics[width=\columnwidth]{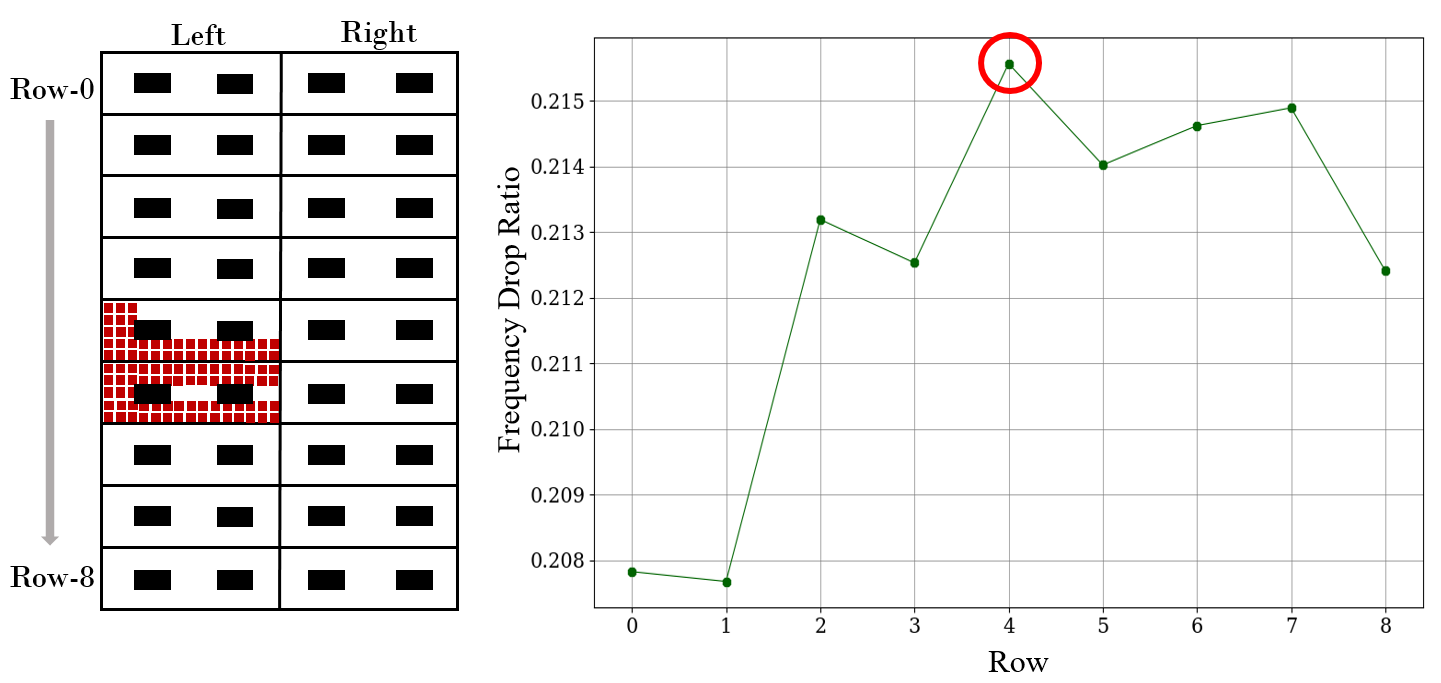}
  \caption{Floorplan and corresponding PRO average frequency drop ratio for each row when power fault happens at location-2. Black blocks donate PROs in the floorplan, red blocks donate power wasters positions in the floorplan.}
  \label{fig:rowplot_loc2}
\end{figure}

\begin{figure}[h]
  \centering
  \includegraphics[width=\columnwidth]{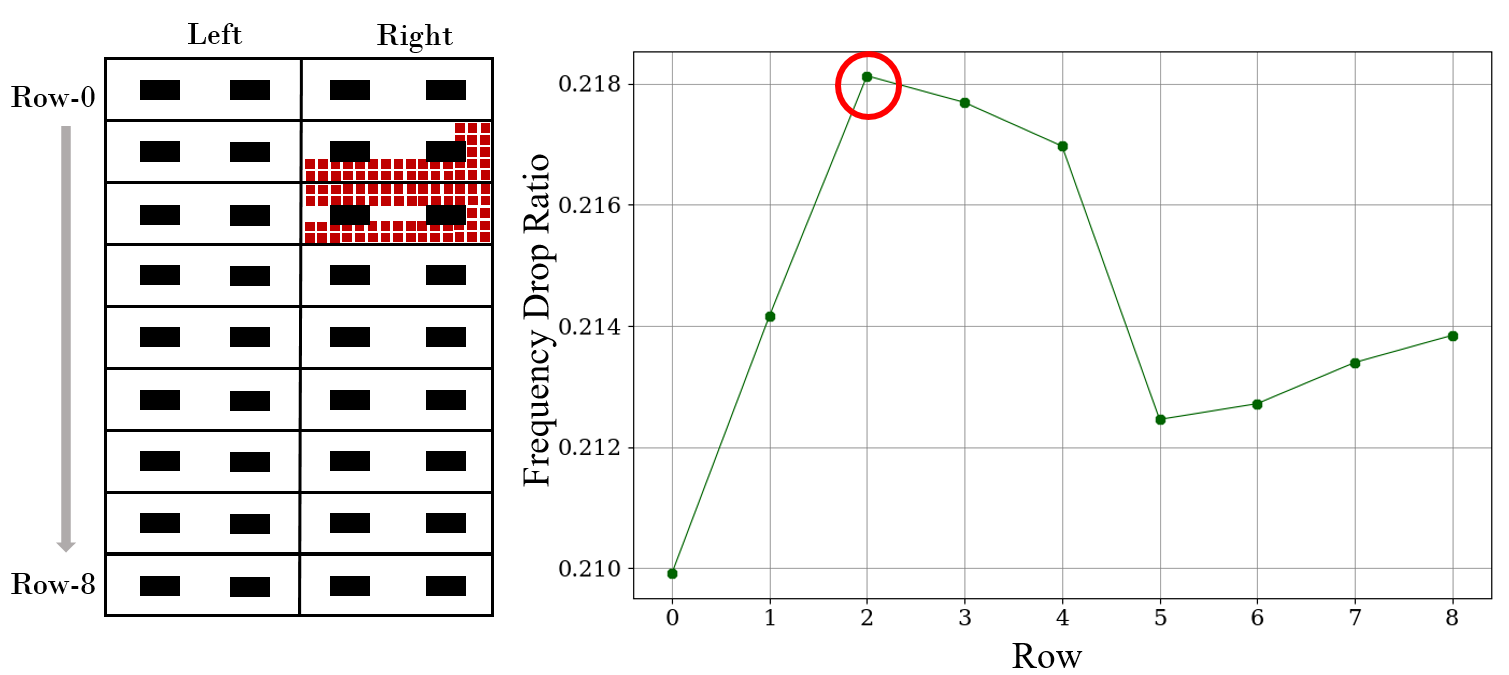}
  \caption{Floorplan and the corresponding PRO average frequency drop ratio for each row when power fault happens at location-3. Black blocks denote PROs in the floorplan, red blocks denote power wasters positions in the floorplan.}
  \label{fig:rowplot_loc3}
\end{figure}

To further demonstrate the capability of the proposed PRO sensor in detecting the location of the fault, we placed the power wasters in different locations to inject fault while AES is running. We repeat the same experimental scenario to locate the faulty row and faulty column. We first put the power wasters in the location Row 4 and Row 5 on the left region and gets the result of locating the faulty row as demonstrated in \autoref{fig:rowplot_loc2}. The highest frequency drop ratio points to Row 4, which indicates that the fault happens adjacent to Row 4. By analyzing the faulty column, we see the left region's average frequency drop is 0.2159 and the right region's frequency drop is 0.2091 which indicates that the left region has the fault. Therefore, the sensor array locates the place where the fault is injected is at the left region Row 4 which meets our expectation. Next, we move the power wasters to another location at Row 1 and Row 2 on the right as shown in \autoref{fig:rowplot_loc3}. We observe the left region's average frequency drop is 0.2083 and the right region's frequency drop is 0.2204 which indicates that the right region has the fault. As shown in the result of analyzing the faulty row in \autoref{fig:rowplot_loc2}, the highest frequency drop ratio points to Row 2 correctly. Therefore, we demonstrate that our proposed sensor can detect the location of the on-chip power fault. 



\subsection{Electromagnetic Fault Injection (EMFI) Detection}

EMFI is a well-known active attack and describes the use of an active probe to apply an intense and transient magnetic field to Integrated Circuits (ICs). EM pulse causes a sudden current flow in the circuit 
of the targeted IC and therefore, the local supply voltage drops. The voltage drop reflects in the form of power consumption peaks. This produces timing faults such as bit-flips, bit-sets, and bit-resets due to timing constraint violation and sampling faults by disrupting the switching process of D-Flip Flops if EM perturbations are synchronous with clock rising edges. This enables the adversary to exploit such faults to extract sensitive content from the device. Previous research has shown that EM perturbations can cause faulty computations, alter the program flow, and cause bit-flips in the contents of the memory. Other authors have demonstrated that EM can induce faults into the devices \cite{8465836, 6076472}. In the past few years, EM fault injection attack has gained increasing attention. In this section, we investigate the performance of our proposed PRO sensor with regard to EM fault injection.   

\begin{figure}[h]
  \centering
  \includegraphics[width=0.85\columnwidth]{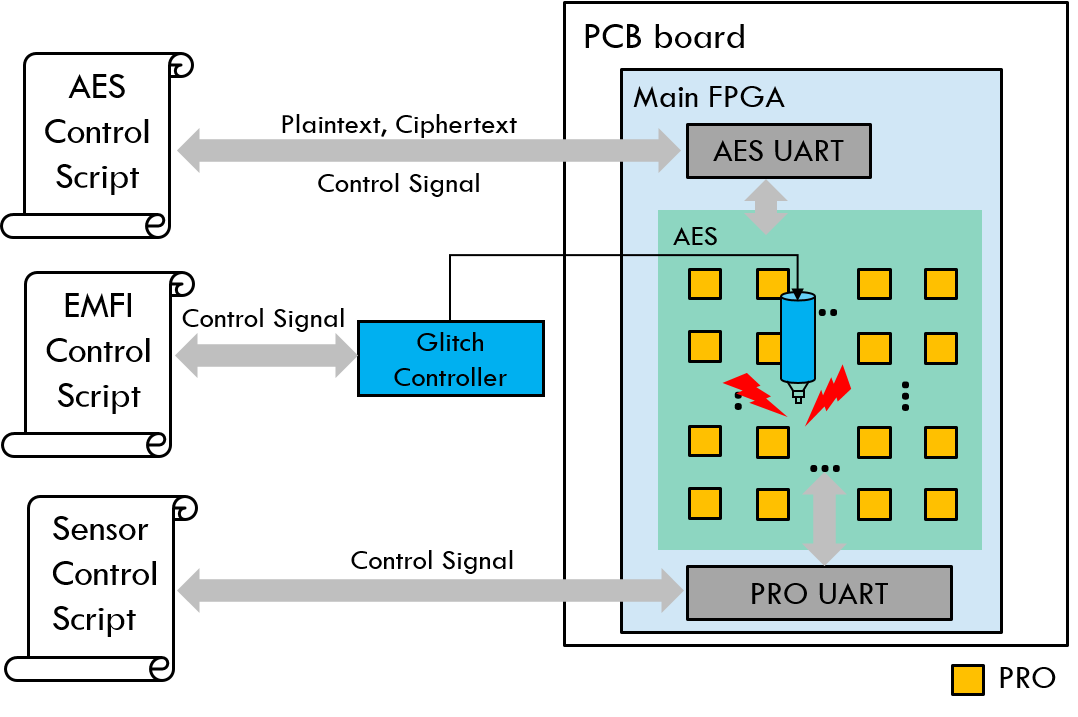}
  \caption{Experimental Setup for PRO EM Fault Detection }
  \label{fig:emfi_setup}
\end{figure}

\autoref{fig:emfi_setup} shows the experimental setup for evaluating the EM fault injection detection performance. In this setup, we instantiate AES and the PRO array with 36 sensors on the FPGA. AES control script is used to control the starting of the AES, send in the plaintext, and read out the ciphertext. While AES is running, the EM probe is placed in a fixed position on top of the FPGA chip surface with a vertical distance of approximately 1.5mm and generates an EM pulse to induce faults. The EM probe's tip is 4mm in diameter and produces a magnetic field that is perpendicular to the surface of the chip. A glitch controller controls the time and intensity of the EM pulse. While AES is running, we adjust the intensity of the EM pulse. When a faulty ciphertext is observed, we know that an effective EM fault is injected. Next, in each measurement, we read out the PRO sensor's counter value through the sensor's control script when the fault is injected and not injected respectively, and compare their values. 

\begin{figure}[h]
  \centering
  \includegraphics[width=\columnwidth]{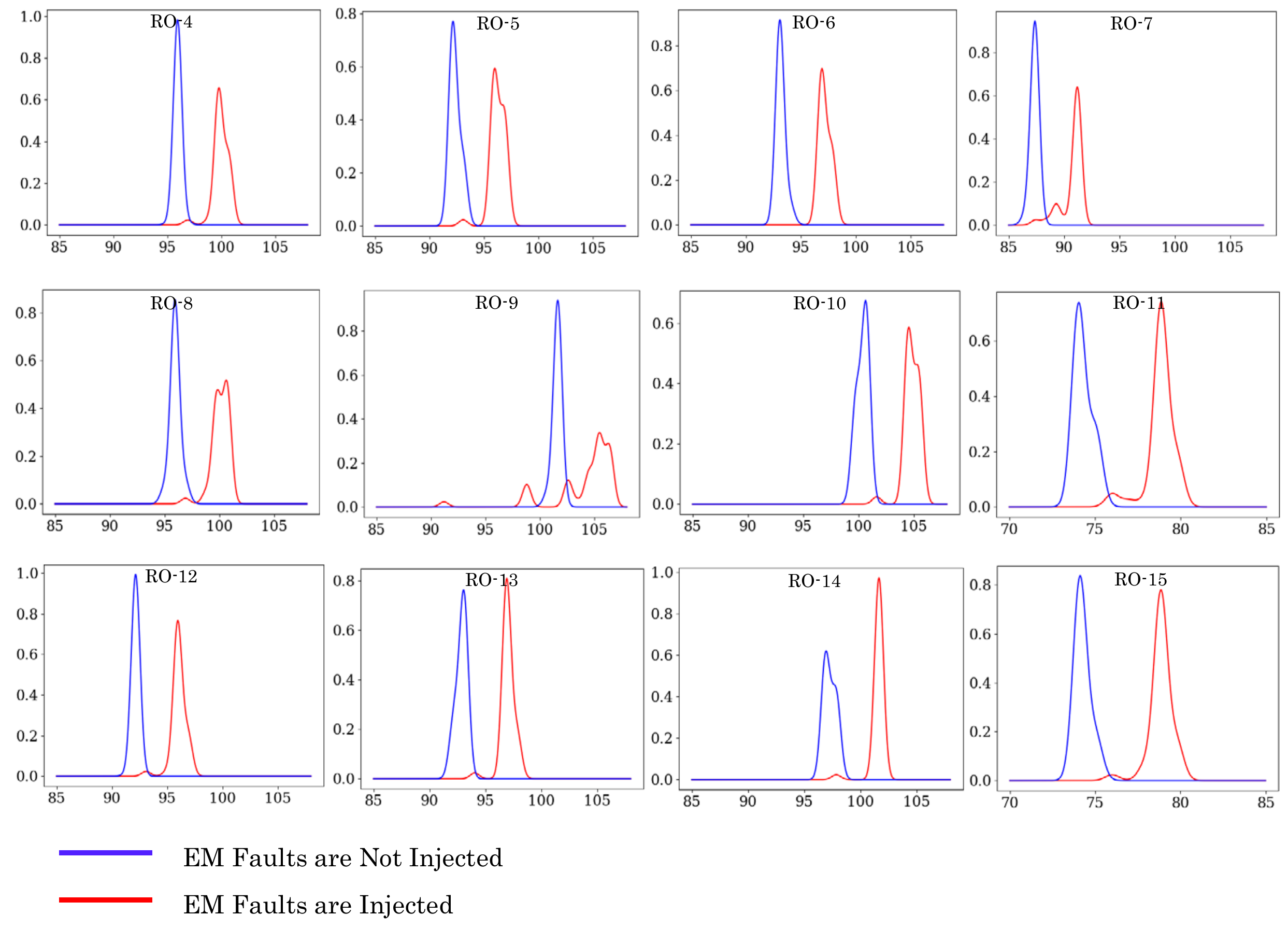}
  \caption{Influence on the Frequency Distribution, X-axis is probability and Y-axis is frequency. }
  \label{fig:EMfault_normal}
\end{figure}

We collect 1000 frequency measurements for all 36 PROs. For each PRO sensor, we investigate the distribution of the 1000 frequency measurements when the EM fault is injected and not injected. We observe that the EM fault can cause variations of the PRO's frequency distribution. \autoref{fig:EMfault_normal} shows comparisons of the frequency distribution when the EM fault is injected and not injected for RO-0 to RO-15. We notice that the PRO sensor's frequency shifts to a larger value when faults are injected. We also observe that besides frequency shifting, there is another fault injection reaction that the PRO sensors can have. We observe that EM faults can also cause faulty counter value for RO-23 to RO-27 and RO-31 to RO-36. When the faults are injected, the counter values are read out by the UART jump to an extremely huge (and faulty) value of  \(4.08\times10^{7}\) MHz.
Therefore, by monitoring the value of the PRO counters, we can detect ongoing \ac{emfi} at run-time.


\begin{figure}[h]
  \centering
  \includegraphics[width=0.8\columnwidth]{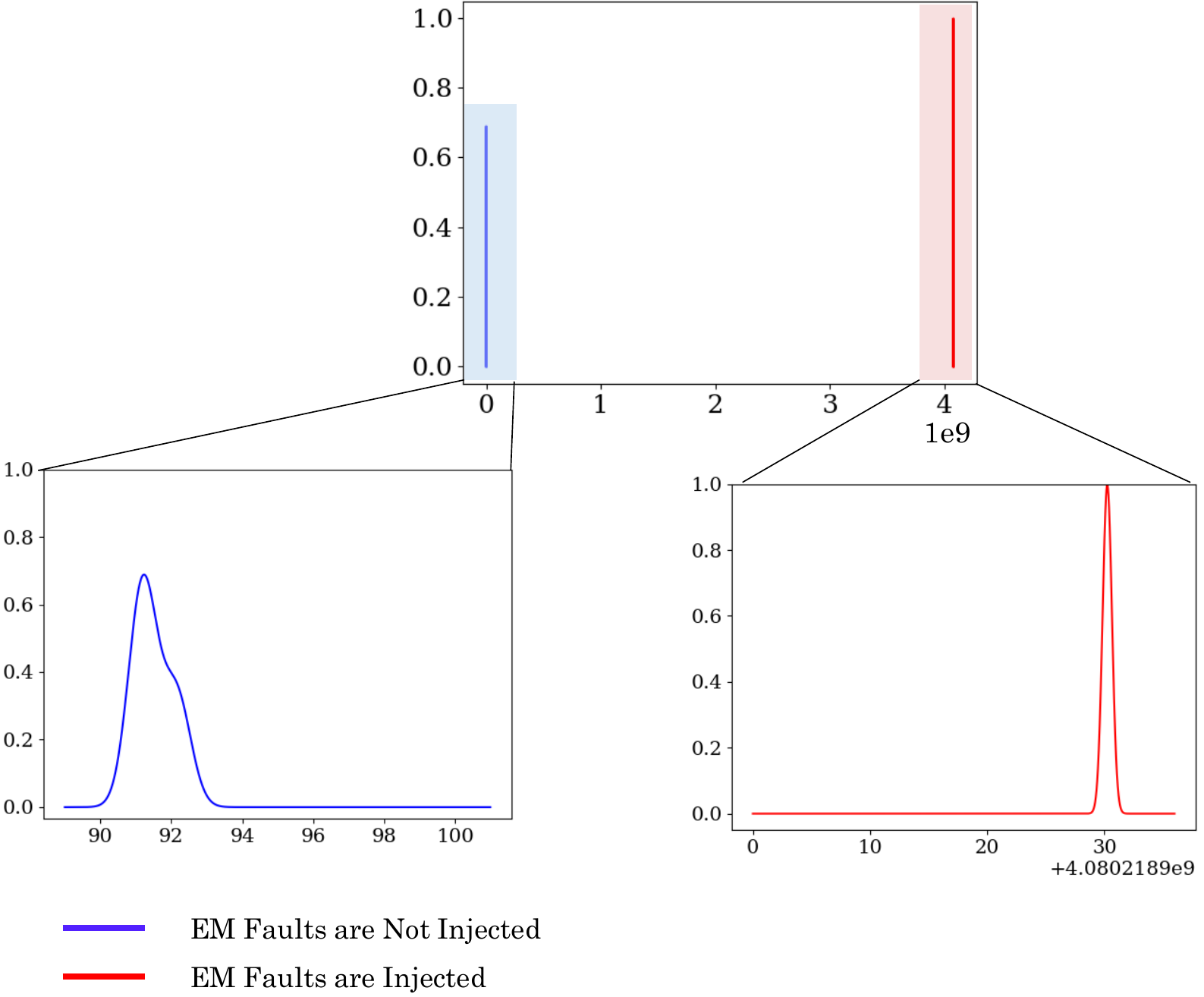}
  \caption{Influence on the Frequency Distribution for PRO-32}
  \label{fig:EMfault_faulty}
\end{figure}

\section{Conclusion} \label{sec:conclusion}

In this work, we proposed a multi-purpose Ring Oscillator design. We demonstrated that it is possible, with a low cost, to have a side-channel countermeasure and fault detection mechanism within the same design. We showed that PRO can provide an effective hiding countermeasure to the circuit with low overhead by injecting random frequency noise. We further demonstrated that the PRO array can form a comprehensive on-chip secure monitoring network. The network can potentially provide both temporal and spatial coverage of on-chip power monitoring and fault detection. PRO has the flexibility for the user to communicate and control its configurations, such as its oscillation frequency, in real-time. This feature highlights its potential to be integrated into large designs, such as SoCs, as a secure extension to build more comprehensive side-channel and fault-resistant systems. As the future work, we will further investigate integrating PRO into an SoC and build up a real-time side-channel countermeasure and fault detection/response system that can protect both software and hardware applications.
\bibliographystyle{ieeepes}
\newlength{\bibitemsep}\setlength{\bibitemsep}{0\baselineskip plus .05\baselineskip minus .05\baselineskip}
\newlength{\bibparskip}\setlength{\bibparskip}{0pt}
\let\oldthebibliography\thebibliography
\renewcommand\thebibliography[1]{%
  \oldthebibliography{#1}%
  \setlength{\parskip}{\bibitemsep}%
  \setlength{\itemsep}{\bibparskip}%
}
\bibliography{ref}

%
\IEEEpeerreviewmaketitle
\vspace*{-1.6cm}
\begin{IEEEbiographynophoto}{Yuan Yao}(Student Member, IEEE) received her bachelor's degree in Electronic Engineering from Northwestern Polytechnical University, Xi'an, China in 2014. She got her Master's Degree in Electrical and Computer Engineering from Cornell University, Ithaca, US in 2016. Currently she is a Ph.D. candidate at the Bradley Department of Electrical and Computer Engineering, Virginia Tech. She serves as reviewer for several IEEE and ACM journals. Her research area include pre-silicon side channel analysis, side channel attacks and countermeasures, fault attacks and countermeasures, secure hardware design. 
\end{IEEEbiographynophoto} 
\vspace*{-1.6cm}
\begin{IEEEbiographynophoto}{Pantea Kiaei} (Student Member, IEEE) is a Ph.D. student in Electrical and Computer Engineering at Worcester Polytechnic Institute. She received her MS degree in Computer Engineering from Virginia Tech in 2019 and prior to that received her BS degree in Electrical Engineering from Sharif University of Technology, Iran, in 2017. She has reviewed papers for ACM TECS and ACM JETC journals. Her research interests include secure hardware design, computer architecture, and trustworthy secure systems.
\end{IEEEbiographynophoto}
\vspace*{-1.6cm}
\begin{IEEEbiographynophoto}{Richa Singh} is a Ph.D. student in Electrical and Computer Engineering at Worcester Polytechnic Institute. She received her B.Tech. degree in Electronics and Communication Engineering from Motilal Nehru National Institute of Technology, India, in 2016. Her research interests include fault attack countermeasures, machine learning and secure hardware design.
\end{IEEEbiographynophoto}
\vspace*{3cm}
\begin{IEEEbiographynophoto}{Shahin Tajik} is an Assistant Professor in Electrical Engineering at Worcester Polytechnic Institute (WPI). Before joining WPI, Dr. Tajik was an Assistant Research Professor at the Florida Institute for Cybersecurity (FICS) Research at the University of Florida. He received his Ph.D. in Electrical Engineering in 2017 from the working group SECT, a collaboration of TU Berlin and Deutsche Telekom Innovation Laboratories in Germany. He received his MS degree in Electrical Engineering in 2013 from TU Berlin in Germany and his BS degree in 2010 from K. N. Toosi University of Technology in Iran.
His field of research primarily includes the physical security evaluation of embedded systems using failure analysis techniques.
\end{IEEEbiographynophoto}
\vspace*{-19cm}
\begin{IEEEbiographynophoto}{Patrick Schaumont} (Senior Member, IEEE) is a Professor in Computer Engineering at WPI. He received the Ph.D. degree in Electrical Engineering from UCLA in 2004 and the MS degree in Computer Science from Ghent University in 1990. He was a staff researcher at IMEC, Belgium from 1992 to 2000. He was a faculty member with Virginia Tech from 2005 to 2019. He joined WPI in 2020. He was a visiting researcher at the National Institute of Information and Telecommunications Technology (NICT), Japan in 2014. He was a visiting researcher at Laboratoire d'Informatique de Paris 6 in Paris, France in 2018. He is a Radboud Excellence Initiative Visiting Faculty with Radboud University, Netherlands from 2020. His research interests are in design and design methods of secure, efficient and real-time embedded computing systems. 
\end{IEEEbiographynophoto}
\vspace*{-1.6cm}











\end{document}